\documentclass[aps,prb,twocolumn,citeautoscript]{revtex4-1}
\usepackage{amsmath,amssymb,mathrsfs,bm,feynmf,setspace}
\usepackage{natbib}
\usepackage{graphicx}
\usepackage[tight]{subfigure}
\usepackage[colorlinks=true]{hyperref} 
\hypersetup{
    bookmarks=true,         
    unicode=false,          
    pdftoolbar=true,        
    pdfmenubar=true,        
    pdffitwindow=false,     
    pdfstartview={FitH},    
    pdftitle={My title},    
    pdfauthor={Author},     
    pdfsubject={Subject},   
    pdfcreator={Creator},   
    pdfproducer={Producer}, 
    pdfkeywords={keyword1} {key2} {key3}, 
    pdfnewwindow=true,      
    colorlinks=true,       
    linkcolor=red,          
    citecolor=blue,        
    filecolor=magenta,      
    urlcolor=cyan           
} 
\usepackage{color}
\newcommand{\al}{\alpha}

\newcommand{\de}{\delta}
\newcommand{\e}{\epsilon}

\newcommand{\s}{\sigma}

\newcommand{\W}{\Omega}

\newcommand{\G}{\Gamma}

\newcommand{\pd}{\partial}

\newcommand{\beq}{\begin{equation}}
\newcommand{\eeq}{\end{equation}}
\newcommand{\Beq}{\begin{eqnarray}}
\newcommand{\Eeq}{\end{eqnarray}}
\newcommand{\bml}{\begin{multline}}

\newcommand{\eeqm}{\end{multline}}

\newcommand{\bsp}{\begin{split}}
\newcommand{\esp}{\end{split}}

\renewcommand{\b}[1]{{\bm #1}}

\newcommand{\inv}{^{-1}}

\newcommand{\mc}{\mathcal}

\newcommand{\req}[1]{Eq.~(\ref{eq:#1})}
\newcommand{\rfig}[1]{Fig.~\ref{fig:#1}}

\newcommand{\nn}{\nonumber}

\begin{document}

\title{Pyrochlore electrons under pressure, heat and field: shedding light on the iridates}
\author{William Witczak-Krempa$^{1}$, Ara Go$^{2}$, Yong Baek Kim$^{1,3}$}
\affiliation{
$^1$Department of Physics, University of Toronto, Toronto, Ontario M5S 1A7, Canada\\
$^2$Department of Physics, Columbia University, New York, New York 10027, USA\\
$^3$School of Physics, Korea Institute for Advanced Study, Seoul 130-722, Korea 
} 
\date{\today}

\begin{abstract}
We study the finite temperature and magnetic field phase diagram of electrons on the
pyrochlore lattice subject to a local repulsion as a model for the pyrochlore iridates.
We provide the most general symmetry-allowed Hamiltonian, including next-nearest neighbour hopping,
and relate it to a Slater-Koster based Hamiltonian for the iridates.
It captures Lifshitz and/or thermal transitions between several phases  
such as metals, semimetals, topological insulators and Weyl semimetals, and 
gapped antiferromagnets with different orders. 
Our results on the charge conductivity, both DC and optical, Hall coefficient, magnetization and 
susceptibility show good agreement with recent experiments and provide new predictions.
As such, our effective model sheds light on the pyrochlore iridates in a unified way.

\end{abstract}
\maketitle 

\section{Introduction}
The pyrochlore iridates, R$_2$Ir$_2$O$_7$ (R-227) where R is a rare earth, 
provide an ideal setup to study the interplay of correlations, band topology and frustration. 
Building in part on the strong role of spin-orbit
coupling (SOC) and correlations for Ir\cite{SrIrO-prl,*SrIrO-science}, various novel phases have been 
predicted to potentially occur in this family of complex oxides
such as fractionalized topological insulators\cite{pesin,will-tmi} (TIs), chiral spin liquids\cite{nakatsuji-pr}, topological Weyl 
semimetals\cite{wan,will-pyro1} (TWSs), and axion insulators\cite{ara-pyro1,wan}.
This has contributed to a substantial experimental 
effort\cite{maeno,taira,maeno-Y-doped,takagi-mit,takagi-mit-full,raman,nakatsuji2011musr,nakatsujiEu,takagi-nd,julian,Nd-pressure,
fisher,disseler-magOrder-nd,disseler-magOrder-y-yb}, 
in particular to determine
the nature of the still-elusive magnetic ordering, except for Pr-227 which shows no sign of long-range order
down to the lowest accessible temperatures. Recent theoretical and experimental work has 
been pointing towards a $q=0$ all-in/all-out (AIO) magnetic pattern in the 
ground state\cite{wan,will-pyro1,nakatsuji2011musr,nakatsujiEu,takagi-nd,disseler-magOrder-nd,disseler-magOrder-y-yb} but direct
evidence is still lacking. Another question concerns the nature of the metal insulator transition as a 
function of chemical pressure via the change of the R-site element\cite{maeno,takagi-mit,takagi-mit-full} or hydrostatic 
pressure\cite{julian,Nd-pressure}, 
as well as the nature of the thermal continuous transitions: 
what are the respective roles of magnetic ordering, Mott physics, and band structure? 
Finally, are any of the new phases, for example the TWS, present in the actual compounds? 

We provide insight into these questions by considering a microscopically-motivated Hubbard model which we probe
at finite temperature and magnetic field. We compute transport, magnetization and susceptibility
in the various phases (metal (M), TI, TWS, AIO, antiferromagnet (AF), etc) as well as their behaviour across the phase transitions, 
quantum or thermal. 
The similarity of our results with recent experimental data allows us to make guesses on the locations 
of different iridates in our phase diagram, and make predictions for future experiments.
Before considering the observables, we discuss a fully general Hubbard Hamiltonian on the pyrochlore 
lattice, its relation to previous models used for the iridates, and to the new model underlying this work. 

The iridium ions form a pyrochlore lattice of corner-sharing tetrahedra and have a 
partially filled shell of 5d electrons. This results in the local repulsion and spin orbit coupling being 
of similar size, both on the order of 0.5-1 eV\cite{pesin,wan}. The localized f-electrons of the R-site (absent for Y) can 
complicate the picture for some members of the iridates family
for which the R-site ion carries a net magnetic moment. In this work we shall not consider the interplay between
these local moments and the more itinerant Ir d-electrons. As such, our results apply directly
to the compounds with Y, Lu and Eu for which this complication does not arise. Regarding the compounds 
with a magnetic R-site, such as for $R=$ Nd, Sm, Yb, etc, 
the Ir d-electron physics might still play a key role. Indeed, the similarity of the phenomena across the pyrochlore iridates
family, independent of whether the R-site is magnetic or not, provide weight to this 
statement\cite{maeno,taira,takagi-mit,takagi-mit-full,raman,disseler-magOrder-y-yb}.
In the context of the pyrochlore iridates, a recent work\cite{chen12} has examined the interplay between the more 
itinerant $d$-electrons and the localized
$f$-electron moments by including an exchange coupling between our nearest neighbor hopping Hamiltonian\cite{will-pyro1}, \req{Hgen}, 
and an appropriate spin model. It was found that Weyl semimetal and axion insulator\cite{wan,ara-pyro1} phases
can be induced by the $f-d$ exchange, and that the latter can cooperate with the Hubbard interaction on Ir sites to stabilize the Weyl semimetal over a larger region of parameter space than when it is induced by $d$-electron correlations alone.
Further studies in that direction will be of interest, notably to get a better understanding 
of the chiral spin liquid behavior of Pr-227\cite{nakatsuji-pr}.

The outline is as follows: we first introduce the general Hamiltonian in Section~\ref{sec:model}
and relate it to previous works. The ground state phase diagram and its extension to finite temperature
is obtained in Section~\ref{sec:pd}. The conductivities, both d.c.\ and optical, are then discussed
in Section~\ref{sec:conductivities}. We discuss the finite-magnetic field phase diagram in Section~\ref{sec:bfield}.
Finally, we briefly summarize in Section~\ref{sec:disc} and draw connections between our results and experiments
on the pyrochlore iridates.  

\section{Model demystified}
\label{sec:model} 
Rather than starting with a specific microscopic model for the iridates,
we consider the most generic, time-reversal invariant Hubbard Hamiltonian on the pyrochlore lattice.
We shall focus on an effective model with a single
Kramers' doublet at each Ir site. 
 This will lead to an 8-band model that we believe is sufficient to capture the salient physics. 
We write the Hamiltonian in a \emph{global} basis for the pseudospin to emphasize its simplicity and generality:
\begin{multline}\label{eq:Hgen}
H_0= \sum_{\substack{
    \langle i, j\rangle }} c_{i}^\dag(t_1+i t_2\b d_{ij}\cdot\b \s) c_{j} \\
+\sum_{\substack{\langle\langle i, j\rangle\rangle }} c_{i}^\dag 
(t_1'+i[t_2' \bm{\mc R}_{ij} + t_3'\bm{\mc D}_{ij}]\cdot\b\s) c_{j}\,,
\end{multline}
where $\b\s$ is a vector of Pauli matrices acting on the pseudospin degree of freedom, 
which in the context of the iridates can be thought of as arising from the splitting of the $t_{2g}$ levels by
SOC\cite{SrIrO-prl,*SrIrO-science} and/or trigonal distortions\cite{bj-trig,fiete-trig} and subsequent 
projection onto a half-filled doublet near the Fermi level ($E_F$).  The real vectors $\b d_{ij}, \bm{\mc R}_{ij},\bm{\mc D}_{ij}$ 
depend on the given bond as follows:
$\b d_{ij}$ is aligned along the opposite bond of the tetrahedron containing $i,j$
(\rfig{H-nn}); it is parallel or antiparallel 
with the nearest neighbor (NN) Dzyaloshinski-Morya (DM) vector of a spin model on the pyrochlore 
lattice, of which there are only two
symmetry-allowed configurations\cite{mc,will-pyro1}, differing by a global sign.
$\bm{\mc R}_{ij}, \bm{\mc D}_{ij}$
are obtained by going to the next nearest neighbour (NNN) via the 
common NN site and taking a cross product of the two
bond or $\b d_{ij}$ vectors encountered, respectively. See Appendix~\ref{ap:hamiltonian} for details
regarding the construction of \req{Hgen}. 
The five real hopping amplitudes lead to a four-parameter free model,
to which we add a Hubbard repulsion term, $H_U=(U/2)\sum_{i} (n_{i}-1)^2$. In this work we perform a mean field
decoupling of $H_U$ in the magnetic channel, but expect many results to survive the inclusion of quantum
many-body effects. Indeed, a subset of many-body effects was taken into account in our recent
study\cite{ara-pyro1} of the NN Hamiltonian using cellular dynamical mean-field theory (CDMFT). For instance, we obtained
the same magnetic orders as those found at the mean-field field level.
\begin{figure}[t]
\centering
\includegraphics[scale=0.75]{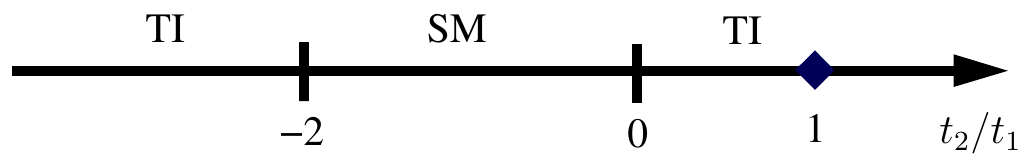}
\caption{\label{fig:t1t2-pd} Phase diagram of the generic NN hopping Hamiltonian 
on a pyrochlore lattice. It has two possible phases: a TI and a SM. 
The square indicates the presence of a gap closing, surrounded by the TI phase. }
\end{figure}

Before discussing the relation of \req{Hgen} to previous microscopic models for the iridates and to the
one used in this work, we briefly discuss the phase diagram including only NN hopping.
It depends on a single parameter, $t_2/t_1$, and can be determined exactly at half-filling; the result is shown in 
\rfig{t1t2-pd}. For $-2<t_2/t_1<0$ a semimetal (SM) results, and otherwise we have a TI\cite{moore-balents,*roy,*fu-kane-mele}, with the exception
of $t_2/t_1=1$ where an accidental gap closing occurs. Interestingly, it does
not alter the topological structure of the Hamiltonian such that the TI survives
to arbitrary large values of $t_2/t_1$. The SM is characterized by lines of 
nodes between the $\G$ and $L$ points.
The phase transitions at $t_2/t_1=-2,0$ occur via gap closing/opening at 
the $\Gamma$ point. At $t_2=0$, the Hamiltonian
can be diagonalized analytically\cite{franz}. Interestingly, it 
is also the case at $t_2/t_1=-2$, where the band structure corresponds exactly to 
the $t_2=0$ case, albeit with a spectrum inversion; specifically: 
$\{t_1=1,t_2=-2\}$ maps to $\{-3,0\}$. The phase transition at $t_2/t_1=0$ was previously discussed\cite{imada},
where the $t_{1,2}$ model was identified as the most general NN symmetry-allowed Hamiltonian. 

We now relate the general parameters to microscopically-motivated ones for the iridates. We
first consider the case of no trigonal distortion\cite{pesin,will-pyro1}, leading to ideal oxygen octahedra surrounding the iridiums.
The SOC splits the $t_{2g}$ manifold into $J_{\rm eff}=1/2$ and $3/2$ multiplets. Projecting onto the half-filled
doublet near $E_F$ yields a pseudospin $J_{\rm eff}=1/2$ description.
Going from the atomic picture to the lattice, we introduce oxygen-mediated\cite{pesin} and direct\cite{will-pyro1} overlap NN hopping 
$t_{\rm oxy}$ and $t_\s,t_\pi$, respectively. These relate to the generic model hoppings via:
\begin{align}
  t_1 &= \frac{130 t_{\rm oxy}}{243}+\frac{17t_\s}{324}  -\frac{79t_\pi}{243}; \quad\,\, t_1' = \frac{233t_\s'}{2916} - \frac{407t_\pi'}{2187}\,;\nn\\
  t_2 &= \frac{28t_{\rm oxy}}{243}+\frac{15 t_\s}{243} -\frac{40 t_\pi}{243}; \,\;\;\;\;\quad  t_2' = \frac{t_\s'}{1458} + \frac{220t_\pi'}{2187}\,;  \nn\\
   t_3' &= \frac{25t_\s'}{1458} + \frac{460t_\pi'}{2187} \,;
\end{align} 
where we have also added $t_{\s,\pi}'$, the $\s$- and $\pi$-overlap NNN hoppings. 
We see that the microscopic model already saturates the general Hamiltonian. As such,
small trigonal distortions\cite{bj-trig} will renormalize the $t_a^{(\prime)}$ but will not contribute new
terms allowing us to use the $J_{\rm eff}=1/2$ description without loss
of generality. 
We mention that Ref.~\onlinecite{pesin} used purely oxygen mediated hopping, hence their large SOC
TI can be mapped to $t_2/t_1=14/65\approx 0.215$ which indeed corresponds to a
TI in \rfig{t1t2-pd}. Also, Ref.~\onlinecite{franz} studied TIs in the $t_1,t_2'$ model.

We shall work in units of $t_{\rm oxy}$ in the rest of the paper.
As before, we choose a representative subset of direct hoppings, $t_\pi=-2t_\s/3$, 
which translates to $t_1=0.53+0.27t_\s$ and $t_2= 0.12 +0.17t_\s$. We plot the relation between $t_2/t_1$ and
$t_\s/t_{\rm oxy}$ in \rfig{t12-vs-ts}.
An important addition in this work is the presence of small NNN hoppings:
$t_\s'/t_\s=t_\pi'/t_\pi=0.08$. These are expected to appear in the effective pseudospin model, and
will remove line Fermi surfaces at $E_F$ in some portions of the phase diagram, giving instead a 
TWS or metal with small pockets\cite{will-pyro1}. The line nodes are an artifact of 
the NN band structure.

\begin{figure*}
    \centering
    \includegraphics[scale=0.65]{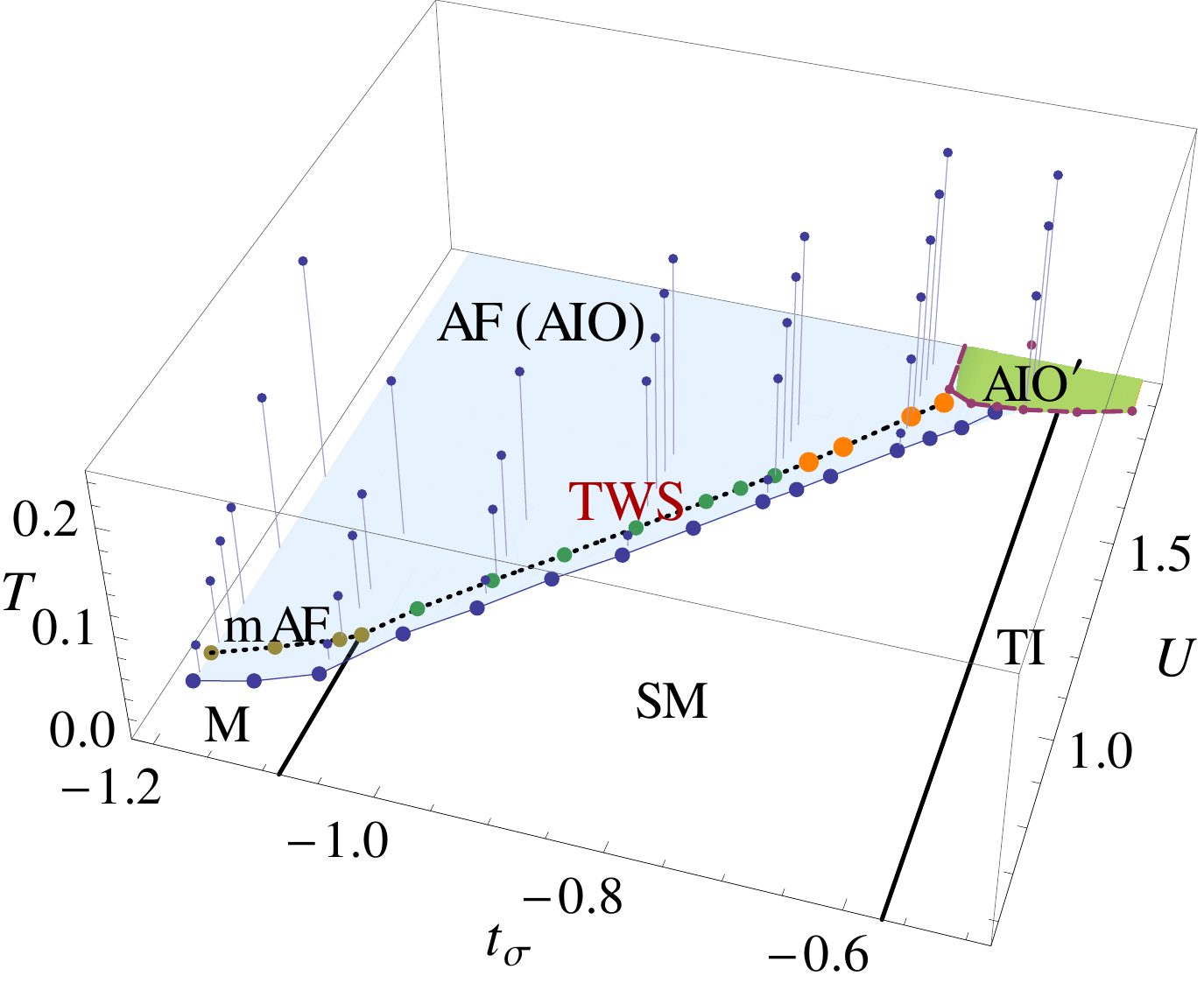}
\caption{\label{fig:pd_nnn008} 
Finite $U$ \& $T$ mean field phase diagram, including small NNN hopping ($t_{\rm oxy}=1$).
The out-of-plane fibers give $T_c$ for the continuous transition at which the magnetic order melts.
The two shaded regions correspond to AF phases: either the
AIO AF, or a type related by $\pi/2$ rotations, AIO$'$.
The solid/dashed lines denote 2nd/1st order quantum phase transitions, while the
dotted lines in the AIO phase, a gap closing signaling a Lifshitz transition out of the
TWS or metallic AF (mAF).  }
\end{figure*}

\section{Phase diagram}
\label{sec:pd}
\begin{figure}
\centering
\subfigure[~$t_\s=-1.2$]{\label{fig:nnn-bs-tsig-12} \includegraphics[scale=.41]{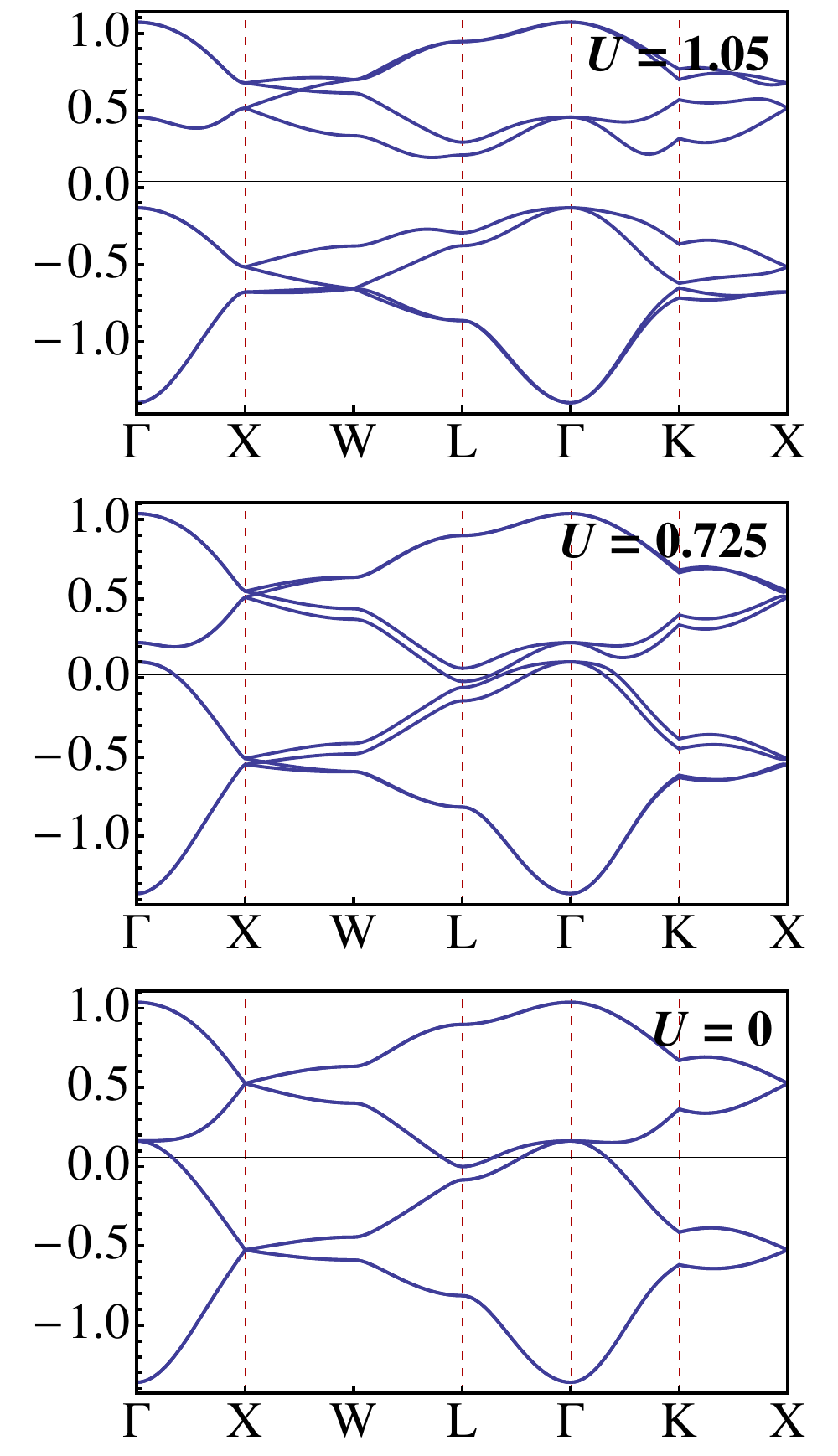}}
\subfigure[~$t_\s=-0.8$]{\label{fig:nnn-bs-tsig-08}\includegraphics[scale=.407]{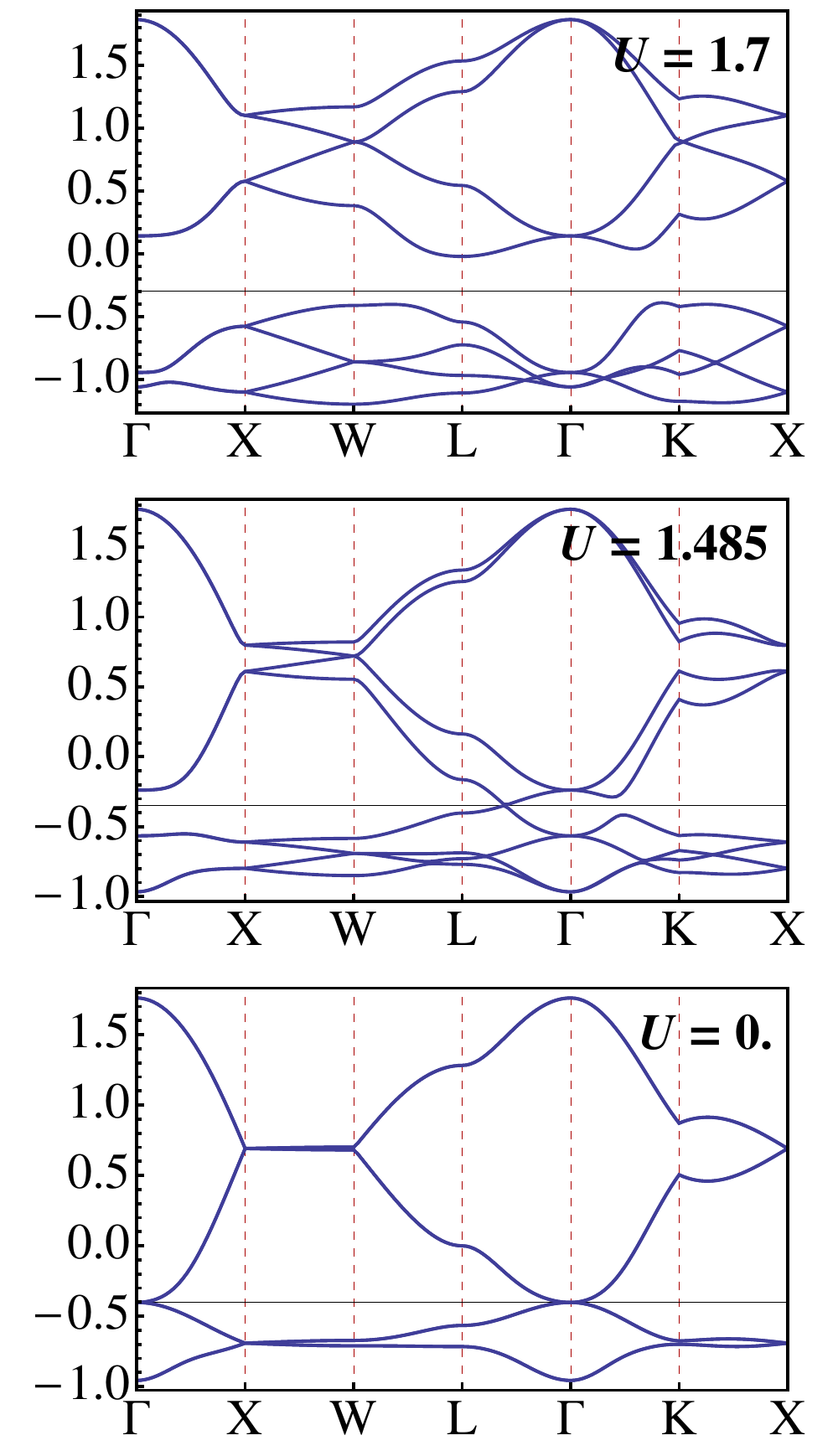}}  
\caption{\label{fig:nnn-bs} 
Spectra for the Hamiltonian \req{Hgen} for (a) $t_\s=-1.2$ and (b) $t_\s=-0.8$.
These are associated with the phase diagram \rfig{pd_nnn008}.}  
\end{figure}

\begin{figure}
    \centering
    \includegraphics[scale=0.4]{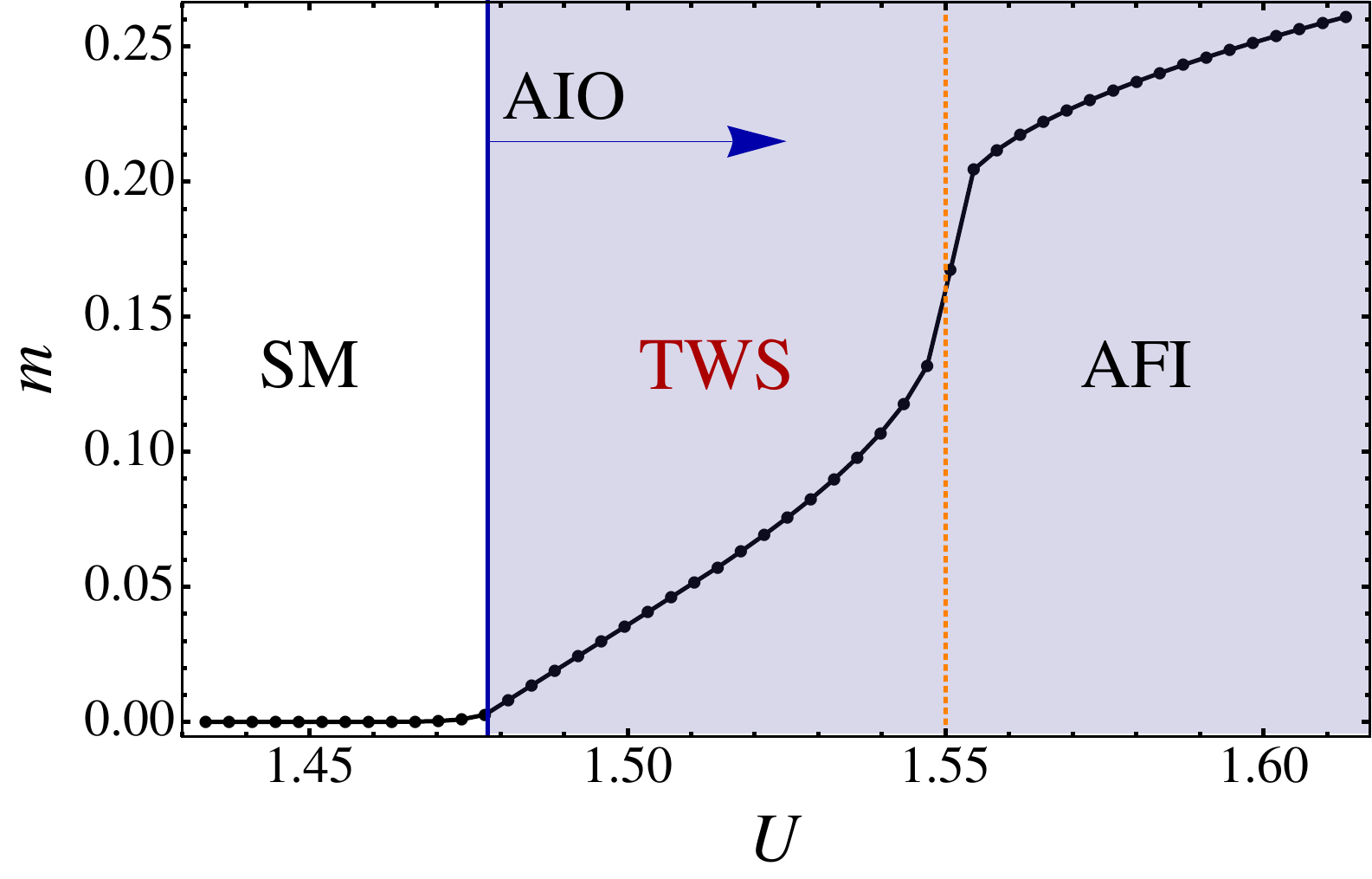}
\caption{\label{fig:jump}  
Evolution of the magnetization, $m$, as a function of the Hubbard repulsion $U$ for $t_\s=-0.775$. Note the abrupt
evolution of $m$ when going from the TWS to the AF insulator (AFI). The shaded regions have the AIO order.}
\end{figure}

The mean field phase diagram is shown in \rfig{pd_nnn008}. At $T=0$ and small $U$, it contains metallic (M), SM and
TI phases depending on $t_\s$. The metallic state has electron- and hole-like pockets while
the SM is associated with a quadratic
band touching, that splits into 8 linearly dispersing touchings in the AF TWS\cite{will-pyro1}, see the
spectra in \rfig{nnn-bs-tsig-08} or in \rfig{opt-sig} c).
The shaded regions break TRS due to the appearance of either the symmetric AIO order (pale blue), or
orders which are related to the AIO state by local $\pi/2$ rotations\cite{will-pyro1} (denoted by AIO$'$, in green).
At intermediate $U$, there is an extended region of TWS, which turns into a metallic AF (mAF) for
$t_\s<-1.085$. The latter is actually a tilted TWS, where the 3D ``Dirac cones'' have been tilted 
such as to create metallic pockets, but with the ``Weyl touchings'' remaining, see \rfig{nnn-bs-tsig-12}.  
These phases should be compared with similar results obtained in Ref.~\cite{will-pyro1} by two of us. The main difference is that
we have now included the NNN hoppings in the self-consistent treatment.
Indeed, the previous work was mainly concerned with the self-consistent analysis of the
NN Hamiltonian: the NNN hoppings were added to the final spectrum to establish that the
line-nodes at $E_F$ can give rise to a Weyl phase when the ordering is of
the AIO type. The present work confirms this. Moreover, a metallic phase now appears for $t_\s<-1.085$, which will be relevant when
we compare with experiments on the iridates. Another new feature is the presence of magnetization
``jumps'' that appear within the AIO phase, present for $-0.78< t_\s < -0.68$, denoted by the orange (larger)
markers in \rfig{pd_nnn008}. The spontaneous onsite magnetization, $m$, grows abruptly along that line
and the TWS becomes a gapped AF insulator: energetically, the system finds it preferable to preempt the continuous annihilation
of Weyl points of opposite chirality via a discontinuous evolution of $m$, this is illustrated in \rfig{jump}.
The abrupt continuous increase of magnetization seems tied to the peculiar spectrum of the TWS.

Turning to finite temperature, we find that most thermal transitions are continuous; this is denoted by the 
out-of-plane fibers ended by circular markers in \rfig{pd_nnn008}.
We have also examined
the model at larger NNN hoppings, $t_{\s,\pi}'/t_{\s,\pi}=0.16$, and found that small regions of 1st order 
transitions to the AIO phase appear at intermediate $U$.
As can be seen in \rfig{pd_nnn008}, the transition temperatures naturally increase with $U$, since an increase in 
the latter leads to larger gaps.
Thus, \emph{the gapless TWS is fairly unstable to melting at finite $T$}, with typical
transition temperatures $T_c\sim 0.01$ in units of $t_{\rm oxy}$. We note that the continuous transitions
are consistent with experimental data on the pyrochlore iridates.


\section{Conductivities}
\label{sec:conductivities}

\subsection{Optical conductivity} 
\begin{figure*}[t]
\centering
\includegraphics[scale=0.9]{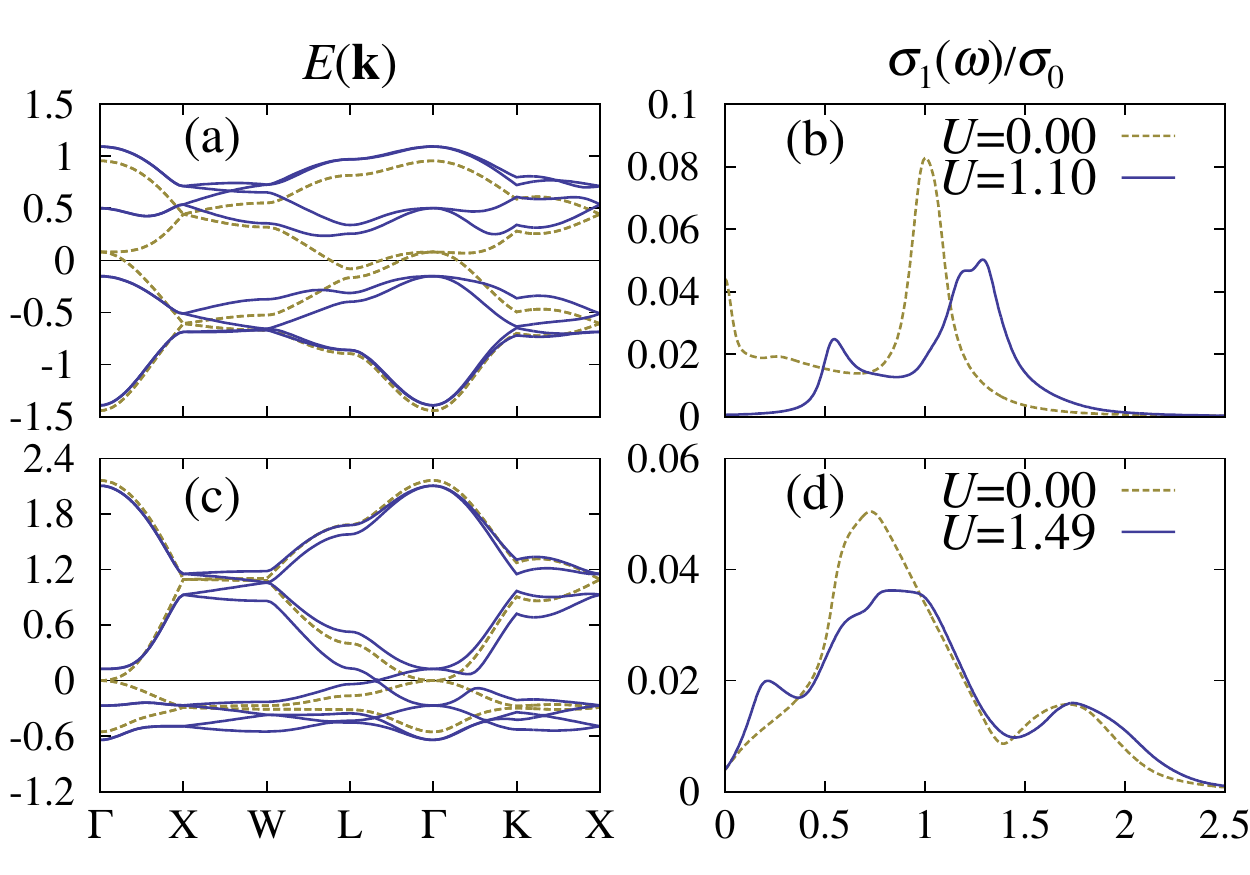}
\caption{\label{fig:opt-sig} 
Spectra and corresponding optical conductivity (real part) for $t_\s=-1.2$, a) \& b), and
$t_\s=-0.8$, c) \& d), respectively. Panel a) shows the energy spectrum ($E_F=0$) for the M and AIO insulator, while
c) for the SM and TWS. The conductivity is in units of $\s_0=e^2/\hbar a_0\sim 10^4\,({\rm \W\cdot cm})\inv$, where $a_0=a/4$ 
and $a\approx 1.0$ nm is the FCC lattice constant.
}
\end{figure*}

We examine the optical conductivity associated with the various phases,
as illustrated in \rfig{opt-sig}. Panel a) shows the Lifshitz transition associated with a change in the Fermi
surface topology from a metal with small electron- and hole-like pockets at $U=0$ to a gapped AIO state at $U=1$. The optical conductivity 
in panel b) reflects this via a small Drude peak in the former case, and a gap in the latter. 
Panel c) shows the 
spectra of the TWS and quadratic band touching SM out of which it arose.
The associated conductivities are reduced approaching the DC limit because these two states have a vanishing 
density of states (DOS) at $E_F$. We note the distinguishing peak at low frequency for the TWS: it receives weight 
from transitions across the depleted energy range corresponding to the TWS, where the DOS vanishes quadratically.

\subsection{DC transport}
\begin{figure}
\centering
\includegraphics[scale=0.55]{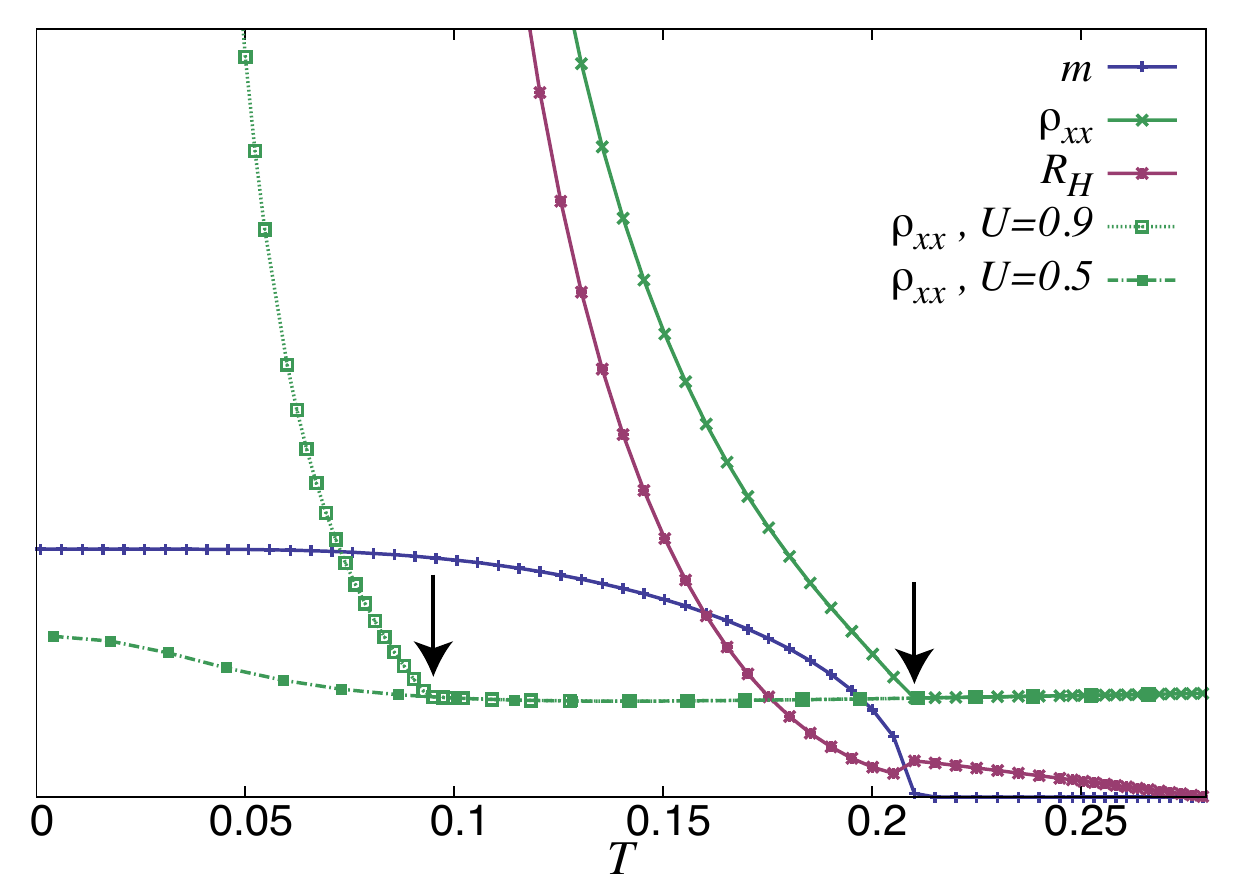}
\caption{\label{fig:transp} 
Solid lines: on-site magnetization ($m$), DC resistivity ($\rho_{xx}$) and Hall coefficient ($R_H$) for $t_\s=-1.2$ and
$U=1.25$. The non-solid lines show $\rho_{xx}$ for $U=0.9$ and $0.5$. The arrows indicate the locations of the continuous 
transitions, at which $\rho_{xx}$ shows a rapid upturn. The vertical axis goes from zero to positive values in arbitrary units.
}
\end{figure}

We turn to the DC electric transport, which provides further insight into the phase diagram. 
The AIO state has isotropic conductivities due to the highly symmetric magnetic configuration, i.e.
$\s_{xx}=\s_{yy}=\s_{zz}$. The same holds for the Hall conductivities: $\s_{xy}=\s_{yz}=\s_{zx}$. This will cease to be true
for the AIO$'$ states, which select a given direction\cite{will-pyro1}. 
To compute the transport coefficients in the DC limit, we introduce a momentum independent scattering time, $\tau$.
For example, the longitudinal and Hall conductivities are given by\cite{ong,daniel}
\begin{align}
  \s_{xx}&=e^2\tau \frac{1}{V}\sum_{\b k,\al}\left(-\frac{\pd n_F}{\pd \e_\al(\b k)}\right)(v^\al_x(\b k))^2 \,, \\
  \s_{xy}&=e^3\tau^2 B_z \frac{1}{V}\sum_{\b k,\al}\left(\frac{\pd n_F}{\pd \e_\al(\b k)}\right)v^\al_y(v^\al_x\pd_y-v^\al_y\pd_x)v^\al_x\,,
\end{align}
where $1\leq\al\leq 8$ is the index associated with the energy band $\e_\al$, the corresponding
band velocity is $\b v^\al=\pd_{\b k}\e_\al$. 
In the Hall conductivity $\s_{xy}$, $B_z$ is the component of the magnetic field in the $z$-direction. 
\rfig{transp} shows the longitudinal resistivity, $\rho_{xx}$, and Hall coefficient, $R_H$, 
at $t_\s=-1.2$ and $U=1.25$ as a function of temperature. At large temperatures, the system is in a paramagnetic metallic state, 
with the expected $d\rho_{xx}/dT>0$. As $T$ is lowered beneath $T_c=0.21$, the system 
undergoes a continuous transition to the AIO AF: the on-site magnetization increases resulting in the spectrum evolving 
from being gapless at $T$ just below $T_c$, to a fully gapped AF at the lowest temperatures. 
We indeed find that the resistivity shows a rapid upturn at $T_c$ (where
the slope changes sign) and becomes exponentially activated at low $T$. The Hall coefficient also shows 
a signature at the transition. Note that the latter is positive suggesting
hole-like carriers in that portion of the phase diagram. The behaviour of the resistivity and sign of the Hall coefficient are
consistent with recent experiments on Eu-227\cite{nakatsujiEu,takagi-mit,takagi-mit-full,julian}. 
\rfig{transp} also shows the evolution of the resistivity as the ratio of the Hubbard repulsion to the bandwidth is reduced. 
The transition occurs at a lower temperature at $U=0.9$ compared to $1.25$, as is expected because the onsite magnetization is reduced
and the corresponding ground state has a smaller gap. At sufficiently small $U$, the system does not develop magnetic ordering and 
the $T=0$ metallic bandstructure is present at all temperatures, in particular this leads to a finite $T=0$ resistivity. 
This behaviour resembles what happens in the iridates as the size of the R-site ion is increased\cite{takagi-mit,takagi-mit-full,maeno} 
(chemical pressure). 
It also roughly agrees with hydrostatic pressure experiments on Eu-227\cite{julian} with the difference that $T_c$ 
was not found to change appreciably with pressure, unlike for Nd-227\cite{Nd-pressure}.

\begin{figure}
\vspace{20pt} 
\centering
\includegraphics[scale=0.5]{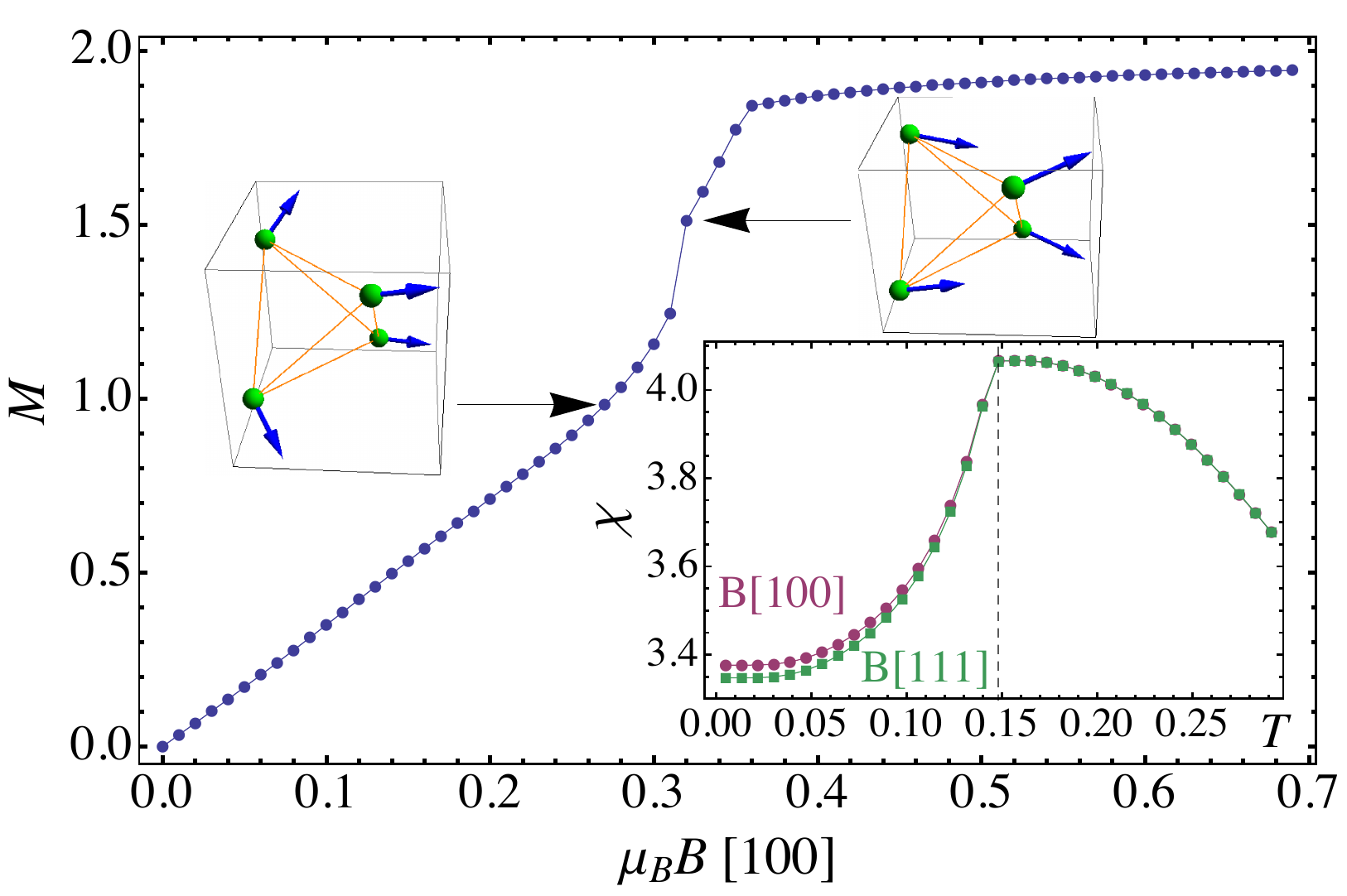}
\caption{\label{fig:mag-chi} 
Main: magnetization per unit cell, $M$, as a function of an applied magnetic field
along the [100] direction for the AIO insulator at $t_\s=-1.2$ and $U=2.1$. The evolution
of the magnetic moments across an abrupt feature is shown. Inset: Susceptibility 
for fields along [100] and [111] versus $T$. The dashed line corresponds to the 2nd 
order transition. 
}
\end{figure}

\section{Magnetic field} 
\label{sec:bfield}
We examine the effects of a magnetic field on some of the ground states of \rfig{pd_nnn008}
via the addition of a Zeeman coupling to the Hubbard Hamiltonian. Since we ignore orbital coupling to the
applied field, we focus on the non-metallic phases. We solve the self-consistent mean-field equations as above
We have considered fields along the [100], [110] and [111] directions.
In all cases, we find a rather continuous evolution to a fully polarized paramagnetic state, without any first order transition to 
a FM aligned with the field, in contrast with claims made in Ref.~\onlinecite{wan}. Instead, the net magnetization grows until it (asymptotically) 
reaches a fully polarized 
state, as shown in \rfig{mag-chi} for the case of the AF insulator at $t_\s=-1.2$ and $U=2.1$. 
The magnetization profiles generically present small features that correspond to sudden changes 
in the magnetic order (never to a FM, though), but lead to small changes in the actual magnetization. One such
abrupt evolution is seen to occur around $\mu_B B\approx 0.32$ in \rfig{mag-chi}.
The magnetic susceptibility, $\chi$, as a function of temperature is shown in the inset of \rfig{mag-chi} for fields along
[100] and [111]; with $\chi[110]=\chi [100]$.
At temperatures above the magnetic melting transition, $T>T_c= 0.148$, $\chi$ is isotropic and slowly varying. 
As $T$ is lowered, $\chi$ has a maximum at $T_c$,
beyond which point it becomes slightly anisotropic. We find that in the ordered phase, $\chi[100]>\chi[111]$.
All of these features are consistent with recent measurements of the zero-field cooled (ZFC) 
susceptibility of Eu-227\cite{nakatsujiEu}.

\section{Discussion} 
\label{sec:disc}
We have introduced a microsopically-motivated Hamiltonian for the pyrochlore iridates, which contains
all symmetry-allowed terms up to NNN. Its ground states include semimetals and metals, topological insulators, topological
Weyl semimetals, gapped AF states with the all-in/all-out and related orders.
Our phase diagram, \rfig{pd_nnn008}, and results for the transport and magnetic properties at finite 
$T$ and field suggest that our new model is apt to describe many salient features of the pyrochlore iridates family. 
In particular, in light of recent experiments\cite{takagi-mit,nakatsujiEu,julian} that show the presence of a charge gap in Eu-227, 
we suggest that its ground state is an AIO gapped insulator, which under pressure can undergo
a Lifshitz transition to a non-magnetic metal with small pockets, although TWS and metallic AF (with coexisting 
Weyl nodes and pockets) are possible. 
At sufficiently high temperatures,
we find that the magnetic order melts continuously leading to a paramagnetic metal, consistent with experiments. 
For Y-227, which shows 
insulating behaviour at all temperatures ($d\rho/dT<0$), we suggest that the high temperature non-magnetic
state (connected to the $U=0$ parent phase) is a semimetal. The TWS, which we find melts at relatively low temperatures, is probably
not present in the ground states of Eu- and Y-227 because of the evidence for a finite gap but it might be 
accessible by hydrostatic pressure\cite{julian,Nd-pressure}. Similarly, the ground state of Nd-227 was suggested\cite{disseler-magOrder-nd}
to be poised near a metal-insulator transition (on the insulating side), potentially in the vicinity of the TWS.

In closing, we mention recent experiments on Bi$_2$Ir$_2$O$_7$\cite{cao-bi227-1,cao-bi227-2} which shows some similar properties to
the other rare-earth based pyrochlore iridates. It was found that the resistivity of Bi-227 remains metallic down to sub-Kelvin 
temperatures\cite{cao-bi227-1}, and recent $\mu$SR measurements\cite{cao-bi227-2} have found evidence for a continuous transition 
into a long-range magnetically ordered state at $\approx 2$ K. Moreover, the measurements suggest that the magnetic moments are 
very small. We observe that these features can be consistently fitted into our framework if we assume the ground state of Bi-227 
is a metallic AF, see \rfig{pd_nnn008}, which is obtained by ``over-tilting'' the TWS. Such a state carries the AIO order with 
small on-site moments. As a consequence, the continuous transition into a paramagnetic state occurs at a 
temperature much smaller than for the phases with a gapped AF ground state. An important caveat is that the Ir $d$-electrons will 
hybridize with 
the $s$- and $p$-orbitals of Bi\cite{cao-bi227-1}, a feature that we do not take into account. 
Nevertheless, since our effective model is quite general, it may not be unreasonable to expect that it will capture some 
features of Bi-227.

\section*{Acknowledgements}
We thank D. Drew and A. Sushkov for sharing preliminary optical conductivity data with us. We acknowledge helpful 
discussions with L.~Balents, P.~Baker,
S.~Bhattacharjee, S. R. Julian, A.J. Millis, S. Nakatsuji, J. Rau, and F. F. Tafti. This work was supported by NSERC, CIFAR, 
the Center for Quantum Materials at the University of Toronto (WWK,YBK), a Walter Sumner fellowship (WWK), and by the
US Department of Energy under grant DOE FG02-04ER46169 (AG). Some part of this work was done at the Aspen Center for Physics, 
partially funded by NSF Grant \#1066293.

\onecolumngrid
\appendix
\section{Constructing the general Hamiltonian}
\label{ap:hamiltonian}

\begin{figure}[h]
\centering%
\subfigure[]{\label{fig:H-nn}\includegraphics[scale=.5]{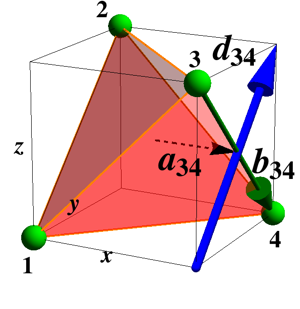}}  
\subfigure[]{\label{fig:H-nnn} \includegraphics[scale=.5]{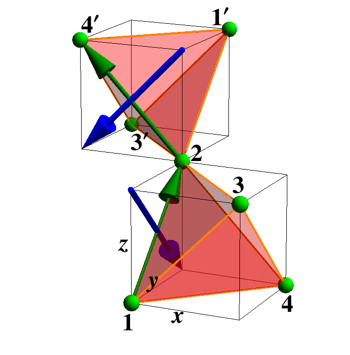}}
\caption{\label{fig:fig-H} Definitions of vectors entering into the general Hamiltonian. The green (thin)
arrows correspond to the $\b b_{ij}$, whereas the blue (thick) ones to the $\b d_{ij}$. b) shows the
vectors involved in obtaining the NNN hopping amplitude between sites 1 and $4'$, who share a common
NN, site 2.
} 
\end{figure}
We provide details regarding the construction of the Hamiltonian used in the main text. 
We choose the basis vectors of the pyrochlore lattice to be \rfig{fig-H} a):
$\b b_1 = (0,0,0)\,, \b b_2 = (0,1,1)\,, \b b_3 = (1,0,1)\,, \b b_4 = (1,1,0)$. In these units
the FCC unit cell has dimension $a=4$. As stated in the main text, the most general 
hopping Hamiltonian on the pyrochlore lattice takes the form:
\begin{align}\label{eq:Hgen-app}
H_0= \sum_{\substack{
    \langle i, j\rangle }} c_{i}^\dag(t_1+i t_2\b d_{ij}\cdot\b \s) c_{j} 
+\sum_{\substack{\langle\langle i, j\rangle\rangle }} c_{i}^\dag 
(t_1'+i[t_2' \bm{\mc R}_{ij} + t_3'\bm{\mc D}_{ij}]\cdot\b\s) c_{j}\,,
\end{align}
where the NN Hamiltonian is built using the vectors:
\begin{align}
  \b d_{ij}&=2 \b a_{ij}\times \b b_{ij}\,, \\
  \b a_{ij}&= \frac{1}{2}(\b b_i+\b b_j) -\b x_c\,,\\
  \b b_{ij}&= \b b_j-\b b_i\,, 
\end{align}
where it is understood in these definitions that $\b b_i$ selects the basis vector of the full lattice (Bravais + basis) index, $i$.
$\b a_{ij}$ points from the center of the tetrahedron, $\b x_c=(1,1,1)/2$, to the middle of the $\langle i,j\rangle$ bond;
it is always orthogonal to the faces of the cube built out of the tetrahedron's vertices, see \rfig{fig-H}. $\b b_{ij}$ is the NN
bond vector pointing from site $i$ to site $j$.
We have chosen the notation for the $\b d_{ij}$ in analogy with the Dzyaloshinski-Morya (DM) vectors, $\b D_{ij}$,
of a spin model on the pyrochlore lattice because both sets of vectors are parallel or antiparallel on all bonds.
The difference arises in the sign. There are only two
symmetry-allowed configurations of DM vectors\cite{mc,will-pyro1}, differing by a global sign. The $\b d_{ij}$'s are always
found to differ by a sign on some bonds with either type of DM vectors.

The NNN Hamiltonian simply uses cross products of the $\b b_{ij}$'s and $\b d_{ij}$'s:
\begin{align}
  \bm{\mc R}_{ij} &=\b b_{ik}\times \b b_{kj}\,, \\
  \bm{\mc D}_{ij} &=\b d_{ik}\times \b d_{kj}\,,
\end{align}
where $i$ and $j$ are NNNs and share the site $k$ as their (unique) common NN.
$\bm{\mc R}_{ij}, \bm{\mc D}_{ij}$ are obtained by going to the NNN via the 
common NN site and taking a cross product of the two bond or $\b d$ vectors encountered, respectively. 
\rfig{fig-H} b) illustrates the vectors necessary to construct the NNN amplitude between sites 1 and $4'$. 
We note that $\bm{\mc R}_{ij}$ and $\bm{\mc D}_{ij}$ always point along the diagonals $(s_1,s_2,s_3)$, where
$s_i=\pm 1$. For instance, $\bm{\mc R}_{34}=(1,-1,1)$ and $\bm{\mc D}_{34}=(-1,1,1)$, as can be
read off from \rfig{fig-H}; they are linearly independent as it should. For completeness, we 
provide the momentum-space form of the Hamiltonian, $H=\sum_{\b q}c_a^\dag(\b q)(\mc H_{ab}^{\rm NN}(\b q)
+\mc H_{ab}^{\rm NNN}(\b q))c_b(\b q)$, with 
\begin{align}
   \mc H_{ab}^{\rm NN}(\b q) &= 2(t_1+t_2i\b\s\cdot \b d_{ab})\cos\left[\b q\cdot\b b_{ab}\right]\,, \\
  \mc H_{ab}^{\rm NNN}(\b q) &= 2\sum_{c\neq a,b}\left\{ t_1'(1-\de_{ab})+i\b\s\cdot\left[t_2'(\b b_{ac}\times\b b_{cb})
      +t_3'(\b d_{ac}\times\b d_{cb})\right] \right\}\cos\left[\b q\cdot (-\b b_{ac}+\b b_{cb})\right]\,,
\end{align}
where the indices $1\leq a,b,c\leq 4$ run over the basis sites. The sum over $c$ in the NNN Hamiltonian 
runs over the choice of common NN between NNNs on sublattices $a$ and $b$. There are two such sites: $c\in\{1,2,3,4\}\setminus \{a,b\}$.

Ref.~\onlinecite{franz} discussed the $t_1,t_2'$ model, with a special focus on topological insulator phases.
Ref.~\onlinecite{imada} identified the $t_1,t_2$ model as the most general NN, time-reversal-invariant hopping 
Hamiltonian on the pyrochlore
lattice, for a single (pseudo)spin-1/2 Hilbert space per site. (We note that 
these two works use $\b d_{ij}$ to denote our $\b b_{ij}$.)

\subsection{Relation to microscopically motivated Hamiltonian}

Two of us have previously proposed a model for the pyrochlore iridates, which was mainly analysed at the
NN level\cite{will-pyro1}. It extends a previous one\cite{pesin} (although we focus on 2 spin/orbital 
degrees of freedom per site instead of 6) to which we added hopping amplitudes
arising from the direct overlaps of the d-orbitals of Ir. These Slater-Koster like models are more naturally
constructed in a local basis because of the relative rotations between the oxygen octahedra within a unit cell,
an illustration of this is found in Ref.~\onlinecite{will-pyro1}. It is however useful to consider the same
Hamiltonians expressed in a global basis instead, to see the role of the different microscopic hoppings more
clearly. 

\begin{figure}
\vspace{20pt} 
\centering
\includegraphics[scale=0.48]{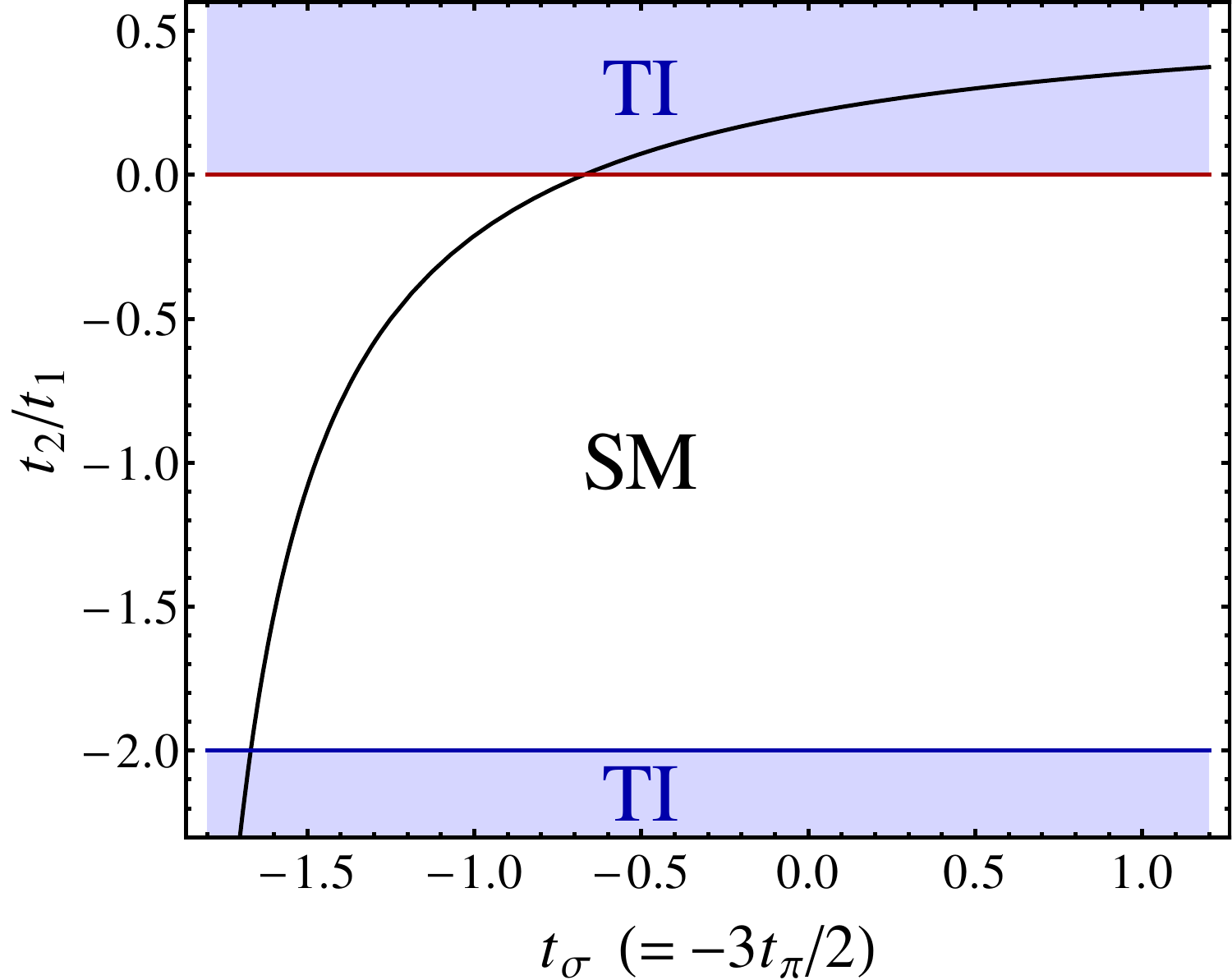}
\caption{\label{fig:t12-vs-ts} 
The curve shows the relation between $t_\s$ (in units of $t_{\rm oxy}$) and $t_2/t_1$, the
latter coming from the general Hamiltonian \req{Hgen}.
}
\end{figure}

We have taken the microscopic, local-axis Hamiltonian given by Eq. (1) in Ref.~\onlinecite{will-pyro1}, supplemented by NNN direct
hoppings, and have performed spinor rotations to obtain a Hamiltonian in a global basis for the pseudospin. This
was compared with \req{Hgen-app} to obtain the relation between $\{t_1,t_2,t_1',t_2',t_3'\}$ and 
$\{t_{\rm oxy},t_\s,t_\pi,t_\s',t_\pi',t_\de'\}$, as given in the main text. Compared with the main body, 
we in addition include the NNN hopping arising from the $\de$-overlap, $t_\de'$, leading to:
\begin{align}
  t_1 &= \frac{130 t_{\rm oxy}}{243}+\frac{17t_\s}{324}  -\frac{79t_\pi}{243}; \quad\,\, 
  t_1' = \frac{233t_\s'}{2916} - \frac{407t_\pi'}{2187}-\frac{1843t_\de'}{8748} \,;\nn\\
  t_2 &= \frac{28t_{\rm oxy}}{243}+\frac{15 t_\s}{243} -\frac{40 t_\pi}{243}; \,\;\;\;\;\quad  
  t_2' = \frac{t_\s'}{1458} + \frac{220t_\pi'}{2187}+ \frac{277t_\de'}{4374}\,;  \nn\\
   t_3' &= \frac{25t_\s'}{1458} + \frac{460t_\pi'}{2187} -\frac{275t_\de'}{4374}\,;
\end{align} 
The NNN hopping $t_\de'$ was added in order to have a number of degrees of freedom 
matching that of the general Hamiltonian at the NNN level. One can also consider the NNN oxygen-mediated
hopping, which will add an analogous term to the $t_i'$. \rfig{t12-vs-ts} shows the relation between 
$t_\s$ (in units of $t_{\rm oxy}$) and $t_2/t_1$, where we have set $t_\pi=-2t_\s/3$. The exact phase boundaries 
of the NN hopping Hamiltonian expressed using the Slater-Koster amplitudes can be obtained from the exact 
ones in the $t_1-t_2$ model, $t_2/t_1=-2,0$:
$t_\s=-192/115, \; -84/125\approx -1.67,\; -0.67$, as can be verified in \rfig{t12-vs-ts}.

\bibliography{ref}{}

\begin{thebibliography}{35}%
\makeatletter
\providecommand \@ifxundefined [1]{%
 \@ifx{#1\undefined}
}%
\providecommand \@ifnum [1]{%
 \ifnum #1\expandafter \@firstoftwo
 \else \expandafter \@secondoftwo
 \fi
}%
\providecommand \@ifx [1]{%
 \ifx #1\expandafter \@firstoftwo
 \else \expandafter \@secondoftwo
 \fi
}%
\providecommand \natexlab [1]{#1}%
\providecommand \enquote  [1]{``#1''}%
\providecommand \bibnamefont  [1]{#1}%
\providecommand \bibfnamefont [1]{#1}%
\providecommand \citenamefont [1]{#1}%
\providecommand \href@noop [0]{\@secondoftwo}%
\providecommand \href [0]{\begingroup \@sanitize@url \@href}%
\providecommand \@href[1]{\@@startlink{#1}\@@href}%
\providecommand \@@href[1]{\endgroup#1\@@endlink}%
\providecommand \@sanitize@url [0]{\catcode `\\12\catcode `\$12\catcode
  `\&12\catcode `\#12\catcode `\^12\catcode `\_12\catcode `\%12\relax}%
\providecommand \@@startlink[1]{}%
\providecommand \@@endlink[0]{}%
\providecommand \url  [0]{\begingroup\@sanitize@url \@url }%
\providecommand \@url [1]{\endgroup\@href {#1}{\urlprefix }}%
\providecommand \urlprefix  [0]{URL }%
\providecommand \Eprint [0]{\href }%
\providecommand \doibase [0]{http://dx.doi.org/}%
\providecommand \selectlanguage [0]{\@gobble}%
\providecommand \bibinfo  [0]{\@secondoftwo}%
\providecommand \bibfield  [0]{\@secondoftwo}%
\providecommand \translation [1]{[#1]}%
\providecommand \BibitemOpen [0]{}%
\providecommand \bibitemStop [0]{}%
\providecommand \bibitemNoStop [0]{.\EOS\space}%
\providecommand \EOS [0]{\spacefactor3000\relax}%
\providecommand \BibitemShut  [1]{\csname bibitem#1\endcsname}%
\let\auto@bib@innerbib\@empty
\bibitem [{\citenamefont {Kim}\ \emph {et~al.}(2008)\citenamefont {Kim},
  \citenamefont {Jin}, \citenamefont {Moon}, \citenamefont {Kim}, \citenamefont
  {Park}, \citenamefont {Leem}, \citenamefont {Yu}, \citenamefont {Noh},
  \citenamefont {Kim}, \citenamefont {Oh}, \citenamefont {Park}, \citenamefont
  {Durairaj}, \citenamefont {Cao},\ and\ \citenamefont
  {Rotenberg}}]{SrIrO-prl}%
  \BibitemOpen
  \bibfield  {author} {\bibinfo {author} {\bibfnamefont {B.~J.}\ \bibnamefont
  {Kim}}, \bibinfo {author} {\bibfnamefont {H.}~\bibnamefont {Jin}}, \bibinfo
  {author} {\bibfnamefont {S.~J.}\ \bibnamefont {Moon}}, \bibinfo {author}
  {\bibfnamefont {J.-Y.}\ \bibnamefont {Kim}}, \bibinfo {author} {\bibfnamefont
  {B.-G.}\ \bibnamefont {Park}}, \bibinfo {author} {\bibfnamefont {C.~S.}\
  \bibnamefont {Leem}}, \bibinfo {author} {\bibfnamefont {J.}~\bibnamefont
  {Yu}}, \bibinfo {author} {\bibfnamefont {T.~W.}\ \bibnamefont {Noh}},
  \bibinfo {author} {\bibfnamefont {C.}~\bibnamefont {Kim}}, \bibinfo {author}
  {\bibfnamefont {S.-J.}\ \bibnamefont {Oh}}, \bibinfo {author} {\bibfnamefont
  {J.-H.}\ \bibnamefont {Park}}, \bibinfo {author} {\bibfnamefont
  {V.}~\bibnamefont {Durairaj}}, \bibinfo {author} {\bibfnamefont
  {G.}~\bibnamefont {Cao}}, \ and\ \bibinfo {author} {\bibfnamefont
  {E.}~\bibnamefont {Rotenberg}},\ }\href {\doibase
  10.1103/PhysRevLett.101.076402} {\bibfield  {journal} {\bibinfo  {journal}
  {Phys. Rev. Lett.}\ }\textbf {\bibinfo {volume} {101}},\ \bibinfo {pages}
  {076402} (\bibinfo {year} {2008})}\BibitemShut {NoStop}%
\bibitem [{\citenamefont {Kim}\ \emph {et~al.}(2009)\citenamefont {Kim},
  \citenamefont {Ohsumi}, \citenamefont {Komesu}, \citenamefont {Sakai},
  \citenamefont {Morita}, \citenamefont {Takagi},\ and\ \citenamefont
  {Arima}}]{SrIrO-science}%
  \BibitemOpen
  \bibfield  {author} {\bibinfo {author} {\bibfnamefont {B.~J.}\ \bibnamefont
  {Kim}}, \bibinfo {author} {\bibfnamefont {H.}~\bibnamefont {Ohsumi}},
  \bibinfo {author} {\bibfnamefont {T.}~\bibnamefont {Komesu}}, \bibinfo
  {author} {\bibfnamefont {S.}~\bibnamefont {Sakai}}, \bibinfo {author}
  {\bibfnamefont {T.}~\bibnamefont {Morita}}, \bibinfo {author} {\bibfnamefont
  {H.}~\bibnamefont {Takagi}}, \ and\ \bibinfo {author} {\bibfnamefont
  {T.}~\bibnamefont {Arima}},\ }\href {\doibase 10.1126/science.1167106}
  {\bibfield  {journal} {\bibinfo  {journal} {Science}\ }\textbf {\bibinfo
  {volume} {323}},\ \bibinfo {pages} {1329} (\bibinfo {year}
  {2009})}\BibitemShut {NoStop}%
\bibitem [{\citenamefont {Pesin}\ and\ \citenamefont {Balents}(2010)}]{pesin}%
  \BibitemOpen
  \bibfield  {author} {\bibinfo {author} {\bibfnamefont {D.}~\bibnamefont
  {Pesin}}\ and\ \bibinfo {author} {\bibfnamefont {L.}~\bibnamefont
  {Balents}},\ }\href {\doibase 10.1038/nphys1606} {\bibfield  {journal}
  {\bibinfo  {journal} {Nat Phys}\ }\textbf {\bibinfo {volume} {6}},\ \bibinfo
  {pages} {376} (\bibinfo {year} {2010})}\BibitemShut {NoStop}%
\bibitem [{\citenamefont {Witczak-Krempa}\ \emph {et~al.}(2010)\citenamefont
  {Witczak-Krempa}, \citenamefont {Choy},\ and\ \citenamefont
  {Kim}}]{will-tmi}%
  \BibitemOpen
  \bibfield  {author} {\bibinfo {author} {\bibfnamefont {W.}~\bibnamefont
  {Witczak-Krempa}}, \bibinfo {author} {\bibfnamefont {T.~P.}\ \bibnamefont
  {Choy}}, \ and\ \bibinfo {author} {\bibfnamefont {Y.~B.}\ \bibnamefont
  {Kim}},\ }\href {\doibase 10.1103/PhysRevB.82.165122} {\bibfield  {journal}
  {\bibinfo  {journal} {Phys. Rev. B}\ }\textbf {\bibinfo {volume} {82}},\
  \bibinfo {pages} {165122} (\bibinfo {year} {2010})}\BibitemShut {NoStop}%
\bibitem [{\citenamefont {Machida}\ \emph {et~al.}(2009)\citenamefont
  {Machida}, \citenamefont {Nakatsuji}, \citenamefont {Onoda}, \citenamefont
  {Tayama},\ and\ \citenamefont {Sakakibara}}]{nakatsuji-pr}%
  \BibitemOpen
  \bibfield  {author} {\bibinfo {author} {\bibfnamefont {Y.}~\bibnamefont
  {Machida}}, \bibinfo {author} {\bibfnamefont {S.}~\bibnamefont {Nakatsuji}},
  \bibinfo {author} {\bibfnamefont {S.}~\bibnamefont {Onoda}}, \bibinfo
  {author} {\bibfnamefont {T.}~\bibnamefont {Tayama}}, \ and\ \bibinfo {author}
  {\bibfnamefont {T.}~\bibnamefont {Sakakibara}},\ }\href {\doibase
  10.1038/nature08680} {\bibfield  {journal} {\bibinfo  {journal} {Nature}\
  }\textbf {\bibinfo {volume} {463}},\ \bibinfo {pages} {210} (\bibinfo {year}
  {2009})}\BibitemShut {NoStop}%
\bibitem [{\citenamefont {Wan}\ \emph {et~al.}(2011)\citenamefont {Wan},
  \citenamefont {Turner}, \citenamefont {Vishwanath},\ and\ \citenamefont
  {Savrasov}}]{wan}%
  \BibitemOpen
  \bibfield  {author} {\bibinfo {author} {\bibfnamefont {X.}~\bibnamefont
  {Wan}}, \bibinfo {author} {\bibfnamefont {A.~M.}\ \bibnamefont {Turner}},
  \bibinfo {author} {\bibfnamefont {A.}~\bibnamefont {Vishwanath}}, \ and\
  \bibinfo {author} {\bibfnamefont {S.~Y.}\ \bibnamefont {Savrasov}},\ }\href
  {\doibase 10.1103/PhysRevB.83.205101} {\bibfield  {journal} {\bibinfo
  {journal} {Phys. Rev. B}\ }\textbf {\bibinfo {volume} {83}},\ \bibinfo
  {pages} {205101} (\bibinfo {year} {2011})}\BibitemShut {NoStop}%
\bibitem [{\citenamefont {Witczak-Krempa}\ and\ \citenamefont
  {Kim}(2012)}]{will-pyro1}%
  \BibitemOpen
  \bibfield  {author} {\bibinfo {author} {\bibfnamefont {W.}~\bibnamefont
  {Witczak-Krempa}}\ and\ \bibinfo {author} {\bibfnamefont {Y.~B.}\
  \bibnamefont {Kim}},\ }\href {\doibase 10.1103/PhysRevB.85.045124} {\bibfield
   {journal} {\bibinfo  {journal} {Phys. Rev. B}\ }\textbf {\bibinfo {volume}
  {85}},\ \bibinfo {pages} {045124} (\bibinfo {year} {2012})}\BibitemShut
  {NoStop}%
\bibitem [{\citenamefont {Go}\ \emph {et~al.}(2012)\citenamefont {Go},
  \citenamefont {Witczak-Krempa}, \citenamefont {Jeon}, \citenamefont {Park},\
  and\ \citenamefont {Kim}}]{ara-pyro1}%
  \BibitemOpen
  \bibfield  {author} {\bibinfo {author} {\bibfnamefont {A.}~\bibnamefont
  {Go}}, \bibinfo {author} {\bibfnamefont {W.}~\bibnamefont {Witczak-Krempa}},
  \bibinfo {author} {\bibfnamefont {G.~S.}\ \bibnamefont {Jeon}}, \bibinfo
  {author} {\bibfnamefont {K.}~\bibnamefont {Park}}, \ and\ \bibinfo {author}
  {\bibfnamefont {Y.~B.}\ \bibnamefont {Kim}},\ }\href {\doibase
  10.1103/PhysRevLett.109.066401} {\bibfield  {journal} {\bibinfo  {journal}
  {Phys. Rev. Lett.}\ }\textbf {\bibinfo {volume} {109}},\ \bibinfo {pages}
  {066401} (\bibinfo {year} {2012})}\BibitemShut {NoStop}%
\bibitem [{\citenamefont {Yanagishima}\ and\ \citenamefont
  {Maeno}(2001)}]{maeno}%
  \BibitemOpen
  \bibfield  {author} {\bibinfo {author} {\bibfnamefont {D.}~\bibnamefont
  {Yanagishima}}\ and\ \bibinfo {author} {\bibfnamefont {Y.}~\bibnamefont
  {Maeno}},\ }\href {\doibase 10.1143/JPSJ.70.2880} {\bibfield  {journal}
  {\bibinfo  {journal} {Journal of the Physical Society of Japan}\ }\textbf
  {\bibinfo {volume} {70}},\ \bibinfo {pages} {2880} (\bibinfo {year}
  {2001})}\BibitemShut {NoStop}%
\bibitem [{\citenamefont {Taira}\ \emph {et~al.}(2001)\citenamefont {Taira},
  \citenamefont {Wakeshima},\ and\ \citenamefont {Hinatsu}}]{taira}%
  \BibitemOpen
  \bibfield  {author} {\bibinfo {author} {\bibfnamefont {N.}~\bibnamefont
  {Taira}}, \bibinfo {author} {\bibfnamefont {M.}~\bibnamefont {Wakeshima}}, \
  and\ \bibinfo {author} {\bibfnamefont {Y.}~\bibnamefont {Hinatsu}},\ }\href
  {\doibase 10.1088/0953-8984/13/23/312} {\bibfield  {journal} {\bibinfo
  {journal} {Journal of Physics: Condensed Matter}\ }\textbf {\bibinfo {volume}
  {13}},\ \bibinfo {pages} {5527+} (\bibinfo {year} {2001})}\BibitemShut
  {NoStop}%
\bibitem [{\citenamefont {Fukazawa}\ and\ \citenamefont
  {Maeno}(2002)}]{maeno-Y-doped}%
  \BibitemOpen
  \bibfield  {author} {\bibinfo {author} {\bibfnamefont {H.}~\bibnamefont
  {Fukazawa}}\ and\ \bibinfo {author} {\bibfnamefont {Y.}~\bibnamefont
  {Maeno}},\ }\href {\doibase 10.1143/JPSJ.71.2578} {\bibfield  {journal}
  {\bibinfo  {journal} {Journal of the Physical Society of Japan}\ }\textbf
  {\bibinfo {volume} {71}},\ \bibinfo {pages} {2578} (\bibinfo {year}
  {2002})}\BibitemShut {NoStop}%
\bibitem [{\citenamefont {Matsuhira}\ \emph {et~al.}(2007)\citenamefont
  {Matsuhira}, \citenamefont {Wakeshima}, \citenamefont {Nakanishi},
  \citenamefont {Yamada}, \citenamefont {Nakamura}, \citenamefont {Kawano},
  \citenamefont {Takagi},\ and\ \citenamefont {Hinatsu}}]{takagi-mit}%
  \BibitemOpen
  \bibfield  {author} {\bibinfo {author} {\bibfnamefont {K.}~\bibnamefont
  {Matsuhira}}, \bibinfo {author} {\bibfnamefont {M.}~\bibnamefont
  {Wakeshima}}, \bibinfo {author} {\bibfnamefont {R.}~\bibnamefont
  {Nakanishi}}, \bibinfo {author} {\bibfnamefont {T.}~\bibnamefont {Yamada}},
  \bibinfo {author} {\bibfnamefont {A.}~\bibnamefont {Nakamura}}, \bibinfo
  {author} {\bibfnamefont {W.}~\bibnamefont {Kawano}}, \bibinfo {author}
  {\bibfnamefont {S.}~\bibnamefont {Takagi}}, \ and\ \bibinfo {author}
  {\bibfnamefont {Y.}~\bibnamefont {Hinatsu}},\ }\href {\doibase
  10.1143/JPSJ.76.043706} {\bibfield  {journal} {\bibinfo  {journal} {Journal
  of the Physical Society of Japan}\ }\textbf {\bibinfo {volume} {76}},\
  \bibinfo {pages} {043706+} (\bibinfo {year} {2007})}\BibitemShut {NoStop}%
\bibitem [{\citenamefont {Matsuhira}\ \emph {et~al.}(2011)\citenamefont
  {Matsuhira}, \citenamefont {Wakeshima}, \citenamefont {Hinatsu},\ and\
  \citenamefont {Takagi}}]{takagi-mit-full}%
  \BibitemOpen
  \bibfield  {author} {\bibinfo {author} {\bibfnamefont {K.}~\bibnamefont
  {Matsuhira}}, \bibinfo {author} {\bibfnamefont {M.}~\bibnamefont
  {Wakeshima}}, \bibinfo {author} {\bibfnamefont {Y.}~\bibnamefont {Hinatsu}},
  \ and\ \bibinfo {author} {\bibfnamefont {S.}~\bibnamefont {Takagi}},\ }\href
  {\doibase 10.1143/JPSJ.80.094701} {\bibfield  {journal} {\bibinfo  {journal}
  {Journal of the Physical Society of Japan}\ }\textbf {\bibinfo {volume}
  {80}},\ \bibinfo {pages} {094701} (\bibinfo {year} {2011})}\BibitemShut
  {NoStop}%
\bibitem [{\citenamefont {Hasegawa}\ \emph {et~al.}(2010)\citenamefont
  {Hasegawa}, \citenamefont {Ogita}, \citenamefont {Matsuhira}, \citenamefont
  {Takagi}, \citenamefont {Wakeshima}, \citenamefont {Hinatsu},\ and\
  \citenamefont {Udagawa}}]{raman}%
  \BibitemOpen
  \bibfield  {author} {\bibinfo {author} {\bibfnamefont {T.}~\bibnamefont
  {Hasegawa}}, \bibinfo {author} {\bibfnamefont {N.}~\bibnamefont {Ogita}},
  \bibinfo {author} {\bibfnamefont {K.}~\bibnamefont {Matsuhira}}, \bibinfo
  {author} {\bibfnamefont {S.}~\bibnamefont {Takagi}}, \bibinfo {author}
  {\bibfnamefont {M.}~\bibnamefont {Wakeshima}}, \bibinfo {author}
  {\bibfnamefont {Y.}~\bibnamefont {Hinatsu}}, \ and\ \bibinfo {author}
  {\bibfnamefont {M.}~\bibnamefont {Udagawa}},\ }\href {\doibase
  10.1088/1742-6596/200/1/012054} {\bibfield  {journal} {\bibinfo  {journal}
  {Journal of Physics: Conference Series}\ }\textbf {\bibinfo {volume} {200}},\
  \bibinfo {pages} {012054+} (\bibinfo {year} {2010})}\BibitemShut {NoStop}%
\bibitem [{\citenamefont {Zhao}\ \emph {et~al.}(2011)\citenamefont {Zhao},
  \citenamefont {Mackie}, \citenamefont {MacLaughlin}, \citenamefont {Bernal},
  \citenamefont {Ishikawa}, \citenamefont {Ohta},\ and\ \citenamefont
  {Nakatsuji}}]{nakatsuji2011musr}%
  \BibitemOpen
  \bibfield  {author} {\bibinfo {author} {\bibfnamefont {S.}~\bibnamefont
  {Zhao}}, \bibinfo {author} {\bibfnamefont {J.~M.}\ \bibnamefont {Mackie}},
  \bibinfo {author} {\bibfnamefont {D.~E.}\ \bibnamefont {MacLaughlin}},
  \bibinfo {author} {\bibfnamefont {O.~O.}\ \bibnamefont {Bernal}}, \bibinfo
  {author} {\bibfnamefont {J.~J.}\ \bibnamefont {Ishikawa}}, \bibinfo {author}
  {\bibfnamefont {Y.}~\bibnamefont {Ohta}}, \ and\ \bibinfo {author}
  {\bibfnamefont {S.}~\bibnamefont {Nakatsuji}},\ }\href {\doibase
  10.1103/PhysRevB.83.180402} {\bibfield  {journal} {\bibinfo  {journal} {Phys.
  Rev. B}\ }\textbf {\bibinfo {volume} {83}},\ \bibinfo {pages} {180402}
  (\bibinfo {year} {2011})}\BibitemShut {NoStop}%
\bibitem [{\citenamefont {Ishikawa}\ \emph {et~al.}(2012)\citenamefont
  {Ishikawa}, \citenamefont {O'Farrell},\ and\ \citenamefont
  {Nakatsuji}}]{nakatsujiEu}%
  \BibitemOpen
  \bibfield  {author} {\bibinfo {author} {\bibfnamefont {J.~J.}\ \bibnamefont
  {Ishikawa}}, \bibinfo {author} {\bibfnamefont {E.~C.~T.}\ \bibnamefont
  {O'Farrell}}, \ and\ \bibinfo {author} {\bibfnamefont {S.}~\bibnamefont
  {Nakatsuji}},\ }\href {\doibase 10.1103/PhysRevB.85.245109} {\bibfield
  {journal} {\bibinfo  {journal} {Phys. Rev. B}\ }\textbf {\bibinfo {volume}
  {85}},\ \bibinfo {pages} {245109} (\bibinfo {year} {2012})}\BibitemShut
  {NoStop}%
\bibitem [{\citenamefont {Tomiyasu}\ \emph {et~al.}(2012)\citenamefont
  {Tomiyasu}, \citenamefont {Matsuhira}, \citenamefont {Iwasa}, \citenamefont
  {Watahiki}, \citenamefont {Takagi}, \citenamefont {Wakeshima}, \citenamefont
  {Hinatsu}, \citenamefont {Yokoyama}, \citenamefont {Ohoyama},\ and\
  \citenamefont {Yamada}}]{takagi-nd}%
  \BibitemOpen
  \bibfield  {author} {\bibinfo {author} {\bibfnamefont {K.}~\bibnamefont
  {Tomiyasu}}, \bibinfo {author} {\bibfnamefont {K.}~\bibnamefont {Matsuhira}},
  \bibinfo {author} {\bibfnamefont {K.}~\bibnamefont {Iwasa}}, \bibinfo
  {author} {\bibfnamefont {M.}~\bibnamefont {Watahiki}}, \bibinfo {author}
  {\bibfnamefont {S.}~\bibnamefont {Takagi}}, \bibinfo {author} {\bibfnamefont
  {M.}~\bibnamefont {Wakeshima}}, \bibinfo {author} {\bibfnamefont
  {Y.}~\bibnamefont {Hinatsu}}, \bibinfo {author} {\bibfnamefont
  {M.}~\bibnamefont {Yokoyama}}, \bibinfo {author} {\bibfnamefont
  {K.}~\bibnamefont {Ohoyama}}, \ and\ \bibinfo {author} {\bibfnamefont
  {K.}~\bibnamefont {Yamada}},\ }\href {\doibase 10.1143/JPSJ.81.034709}
  {\bibfield  {journal} {\bibinfo  {journal} {Journal of the Physical Society
  of Japan}\ }\textbf {\bibinfo {volume} {81}},\ \bibinfo {pages} {034709+}
  (\bibinfo {year} {2012})}\BibitemShut {NoStop}%
\bibitem [{\citenamefont {Tafti}\ \emph {et~al.}(2012)\citenamefont {Tafti},
  \citenamefont {Ishikawa}, \citenamefont {McCollam}, \citenamefont
  {Nakatsuji},\ and\ \citenamefont {Julian}}]{julian}%
  \BibitemOpen
  \bibfield  {author} {\bibinfo {author} {\bibfnamefont {F.~F.}\ \bibnamefont
  {Tafti}}, \bibinfo {author} {\bibfnamefont {J.~J.}\ \bibnamefont {Ishikawa}},
  \bibinfo {author} {\bibfnamefont {A.}~\bibnamefont {McCollam}}, \bibinfo
  {author} {\bibfnamefont {S.}~\bibnamefont {Nakatsuji}}, \ and\ \bibinfo
  {author} {\bibfnamefont {S.~R.}\ \bibnamefont {Julian}},\ }\href {\doibase
  10.1103/PhysRevB.85.205104} {\bibfield  {journal} {\bibinfo  {journal} {Phys.
  Rev. B}\ }\textbf {\bibinfo {volume} {85}},\ \bibinfo {pages} {205104}
  (\bibinfo {year} {2012})}\BibitemShut {NoStop}%
\bibitem [{\citenamefont {Sakata}\ \emph {et~al.}(2011)\citenamefont {Sakata},
  \citenamefont {Kagayama}, \citenamefont {Shimizu}, \citenamefont {Matsuhira},
  \citenamefont {Takagi}, \citenamefont {Wakeshima},\ and\ \citenamefont
  {Hinatsu}}]{Nd-pressure}%
  \BibitemOpen
  \bibfield  {author} {\bibinfo {author} {\bibfnamefont {M.}~\bibnamefont
  {Sakata}}, \bibinfo {author} {\bibfnamefont {T.}~\bibnamefont {Kagayama}},
  \bibinfo {author} {\bibfnamefont {K.}~\bibnamefont {Shimizu}}, \bibinfo
  {author} {\bibfnamefont {K.}~\bibnamefont {Matsuhira}}, \bibinfo {author}
  {\bibfnamefont {S.}~\bibnamefont {Takagi}}, \bibinfo {author} {\bibfnamefont
  {M.}~\bibnamefont {Wakeshima}}, \ and\ \bibinfo {author} {\bibfnamefont
  {Y.}~\bibnamefont {Hinatsu}},\ }\href {\doibase 10.1103/PhysRevB.83.041102}
  {\bibfield  {journal} {\bibinfo  {journal} {Phys. Rev. B}\ }\textbf {\bibinfo
  {volume} {83}},\ \bibinfo {pages} {041102} (\bibinfo {year}
  {2011})}\BibitemShut {NoStop}%
\bibitem [{\citenamefont {Shapiro}\ \emph {et~al.}(2012)\citenamefont
  {Shapiro}, \citenamefont {Riggs}, \citenamefont {Stone}, \citenamefont {de~la
  Cruz}, \citenamefont {Chi}, \citenamefont {Podlesnyak},\ and\ \citenamefont
  {Fisher}}]{fisher}%
  \BibitemOpen
  \bibfield  {author} {\bibinfo {author} {\bibfnamefont {M.~C.}\ \bibnamefont
  {Shapiro}}, \bibinfo {author} {\bibfnamefont {S.~C.}\ \bibnamefont {Riggs}},
  \bibinfo {author} {\bibfnamefont {M.~B.}\ \bibnamefont {Stone}}, \bibinfo
  {author} {\bibfnamefont {C.~R.}\ \bibnamefont {de~la Cruz}}, \bibinfo
  {author} {\bibfnamefont {S.}~\bibnamefont {Chi}}, \bibinfo {author}
  {\bibfnamefont {A.~A.}\ \bibnamefont {Podlesnyak}}, \ and\ \bibinfo {author}
  {\bibfnamefont {I.~R.}\ \bibnamefont {Fisher}},\ }\href {\doibase
  10.1103/PhysRevB.85.214434} {\bibfield  {journal} {\bibinfo  {journal} {Phys.
  Rev. B}\ }\textbf {\bibinfo {volume} {85}},\ \bibinfo {pages} {214434}
  (\bibinfo {year} {2012})}\BibitemShut {NoStop}%
\bibitem [{\citenamefont {Disseler}\ \emph
  {et~al.}(2012{\natexlab{a}})\citenamefont {Disseler}, \citenamefont {Dhital},
  \citenamefont {Hogan}, \citenamefont {Amato}, \citenamefont {Giblin},
  \citenamefont {de~la Cruz}, \citenamefont {Daoud-Aladine}, \citenamefont
  {Wilson},\ and\ \citenamefont {Graf}}]{disseler-magOrder-nd}%
  \BibitemOpen
  \bibfield  {author} {\bibinfo {author} {\bibfnamefont {S.~M.}\ \bibnamefont
  {Disseler}}, \bibinfo {author} {\bibfnamefont {C.}~\bibnamefont {Dhital}},
  \bibinfo {author} {\bibfnamefont {T.~C.}\ \bibnamefont {Hogan}}, \bibinfo
  {author} {\bibfnamefont {A.}~\bibnamefont {Amato}}, \bibinfo {author}
  {\bibfnamefont {S.~R.}\ \bibnamefont {Giblin}}, \bibinfo {author}
  {\bibfnamefont {C.}~\bibnamefont {de~la Cruz}}, \bibinfo {author}
  {\bibfnamefont {A.}~\bibnamefont {Daoud-Aladine}}, \bibinfo {author}
  {\bibfnamefont {S.~D.}\ \bibnamefont {Wilson}}, \ and\ \bibinfo {author}
  {\bibfnamefont {M.~J.}\ \bibnamefont {Graf}},\ }\href {\doibase
  10.1103/PhysRevB.85.174441} {\bibfield  {journal} {\bibinfo  {journal} {Phys.
  Rev. B}\ }\textbf {\bibinfo {volume} {85}},\ \bibinfo {pages} {174441}
  (\bibinfo {year} {2012}{\natexlab{a}})}\BibitemShut {NoStop}%
\bibitem [{\citenamefont {Disseler}\ \emph
  {et~al.}(2012{\natexlab{b}})\citenamefont {Disseler}, \citenamefont {Dhital},
  \citenamefont {Amato}, \citenamefont {Giblin}, \citenamefont {de~la Cruz},
  \citenamefont {Wilson},\ and\ \citenamefont {Graf}}]{disseler-magOrder-y-yb}%
  \BibitemOpen
  \bibfield  {author} {\bibinfo {author} {\bibfnamefont {S.~M.}\ \bibnamefont
  {Disseler}}, \bibinfo {author} {\bibfnamefont {C.}~\bibnamefont {Dhital}},
  \bibinfo {author} {\bibfnamefont {A.}~\bibnamefont {Amato}}, \bibinfo
  {author} {\bibfnamefont {S.~R.}\ \bibnamefont {Giblin}}, \bibinfo {author}
  {\bibfnamefont {C.}~\bibnamefont {de~la Cruz}}, \bibinfo {author}
  {\bibfnamefont {S.~D.}\ \bibnamefont {Wilson}}, \ and\ \bibinfo {author}
  {\bibfnamefont {M.~J.}\ \bibnamefont {Graf}},\ }\href {\doibase
  10.1103/PhysRevB.86.014428} {\bibfield  {journal} {\bibinfo  {journal} {Phys.
  Rev. B}\ }\textbf {\bibinfo {volume} {86}},\ \bibinfo {pages} {014428}
  (\bibinfo {year} {2012}{\natexlab{b}})}\BibitemShut {NoStop}%
\bibitem [{\citenamefont {Chen}\ and\ \citenamefont {Hermele}(2012)}]{chen12}%
  \BibitemOpen
  \bibfield  {author} {\bibinfo {author} {\bibfnamefont {G.}~\bibnamefont
  {Chen}}\ and\ \bibinfo {author} {\bibfnamefont {M.}~\bibnamefont {Hermele}},\
  }\href {\doibase 10.1103/PhysRevB.86.235129} {\bibfield  {journal} {\bibinfo
  {journal} {Phys. Rev. B}\ }\textbf {\bibinfo {volume} {86}},\ \bibinfo
  {pages} {235129} (\bibinfo {year} {2012})}\BibitemShut {NoStop}%
\bibitem [{\citenamefont {Yang}\ and\ \citenamefont {Kim}(2010)}]{bj-trig}%
  \BibitemOpen
  \bibfield  {author} {\bibinfo {author} {\bibfnamefont {B.-J.}\ \bibnamefont
  {Yang}}\ and\ \bibinfo {author} {\bibfnamefont {Y.~B.}\ \bibnamefont {Kim}},\
  }\href {\doibase 10.1103/PhysRevB.82.085111} {\bibfield  {journal} {\bibinfo
  {journal} {Phys. Rev. B}\ }\textbf {\bibinfo {volume} {82}},\ \bibinfo
  {pages} {085111} (\bibinfo {year} {2010})}\BibitemShut {NoStop}%
\bibitem [{\citenamefont {Kargarian}\ \emph {et~al.}(2011)\citenamefont
  {Kargarian}, \citenamefont {Wen},\ and\ \citenamefont {Fiete}}]{fiete-trig}%
  \BibitemOpen
  \bibfield  {author} {\bibinfo {author} {\bibfnamefont {M.}~\bibnamefont
  {Kargarian}}, \bibinfo {author} {\bibfnamefont {J.}~\bibnamefont {Wen}}, \
  and\ \bibinfo {author} {\bibfnamefont {G.~A.}\ \bibnamefont {Fiete}},\ }\href
  {\doibase 10.1103/PhysRevB.83.165112} {\bibfield  {journal} {\bibinfo
  {journal} {Phys. Rev. B}\ }\textbf {\bibinfo {volume} {83}},\ \bibinfo
  {pages} {165112} (\bibinfo {year} {2011})}\BibitemShut {NoStop}%
\bibitem [{\citenamefont {Elhajal}\ \emph {et~al.}(2005)\citenamefont
  {Elhajal}, \citenamefont {Canals}, \citenamefont {Sunyer},\ and\
  \citenamefont {Lacroix}}]{mc}%
  \BibitemOpen
  \bibfield  {author} {\bibinfo {author} {\bibfnamefont {M.}~\bibnamefont
  {Elhajal}}, \bibinfo {author} {\bibfnamefont {B.}~\bibnamefont {Canals}},
  \bibinfo {author} {\bibfnamefont {R.}~\bibnamefont {Sunyer}}, \ and\ \bibinfo
  {author} {\bibfnamefont {C.}~\bibnamefont {Lacroix}},\ }\href@noop {}
  {\bibfield  {journal} {\bibinfo  {journal} {Phys. Rev. B}\ }\textbf {\bibinfo
  {volume} {71}},\ \bibinfo {pages} {094420} (\bibinfo {year}
  {2005})}\BibitemShut {NoStop}%
\bibitem [{\citenamefont {Moore}\ and\ \citenamefont
  {Balents}(2007)}]{moore-balents}%
  \BibitemOpen
  \bibfield  {author} {\bibinfo {author} {\bibfnamefont {J.~E.}\ \bibnamefont
  {Moore}}\ and\ \bibinfo {author} {\bibfnamefont {L.}~\bibnamefont
  {Balents}},\ }\href {\doibase 10.1103/PhysRevB.75.121306} {\bibfield
  {journal} {\bibinfo  {journal} {Phys. Rev. B}\ }\textbf {\bibinfo {volume}
  {75}},\ \bibinfo {pages} {121306} (\bibinfo {year} {2007})}\BibitemShut
  {NoStop}%
\bibitem [{\citenamefont {Roy}(2009)}]{roy}%
  \BibitemOpen
  \bibfield  {author} {\bibinfo {author} {\bibfnamefont {R.}~\bibnamefont
  {Roy}},\ }\href {\doibase 10.1103/PhysRevB.79.195322} {\bibfield  {journal}
  {\bibinfo  {journal} {Phys. Rev. B}\ }\textbf {\bibinfo {volume} {79}},\
  \bibinfo {pages} {195322} (\bibinfo {year} {2009})}\BibitemShut {NoStop}%
\bibitem [{\citenamefont {Fu}\ \emph {et~al.}(2007)\citenamefont {Fu},
  \citenamefont {Kane},\ and\ \citenamefont {Mele}}]{fu-kane-mele}%
  \BibitemOpen
  \bibfield  {author} {\bibinfo {author} {\bibfnamefont {L.}~\bibnamefont
  {Fu}}, \bibinfo {author} {\bibfnamefont {C.~L.}\ \bibnamefont {Kane}}, \ and\
  \bibinfo {author} {\bibfnamefont {E.~J.}\ \bibnamefont {Mele}},\ }\href
  {\doibase 10.1103/PhysRevLett.98.106803} {\bibfield  {journal} {\bibinfo
  {journal} {Phys. Rev. Lett.}\ }\textbf {\bibinfo {volume} {98}},\ \bibinfo
  {pages} {106803} (\bibinfo {year} {2007})}\BibitemShut {NoStop}%
\bibitem [{\citenamefont {Guo}\ and\ \citenamefont {Franz}(2009)}]{franz}%
  \BibitemOpen
  \bibfield  {author} {\bibinfo {author} {\bibfnamefont {H.-M.}\ \bibnamefont
  {Guo}}\ and\ \bibinfo {author} {\bibfnamefont {M.}~\bibnamefont {Franz}},\
  }\href {\doibase 10.1103/PhysRevLett.103.206805} {\bibfield  {journal}
  {\bibinfo  {journal} {Phys. Rev. Lett.}\ }\textbf {\bibinfo {volume} {103}},\
  \bibinfo {pages} {206805} (\bibinfo {year} {2009})}\BibitemShut {NoStop}%
\bibitem [{\citenamefont {Kurita}\ \emph {et~al.}(2011)\citenamefont {Kurita},
  \citenamefont {Yamaji},\ and\ \citenamefont {Imada}}]{imada}%
  \BibitemOpen
  \bibfield  {author} {\bibinfo {author} {\bibfnamefont {M.}~\bibnamefont
  {Kurita}}, \bibinfo {author} {\bibfnamefont {Y.}~\bibnamefont {Yamaji}}, \
  and\ \bibinfo {author} {\bibfnamefont {M.}~\bibnamefont {Imada}},\ }\href
  {\doibase 10.1143/JPSJ.80.044708} {\bibfield  {journal} {\bibinfo  {journal}
  {Journal of the Physical Society of Japan}\ }\textbf {\bibinfo {volume}
  {80}},\ \bibinfo {pages} {044708+} (\bibinfo {year} {2011})}\BibitemShut
  {NoStop}%
\bibitem [{\citenamefont {Ong}(1991)}]{ong}%
  \BibitemOpen
  \bibfield  {author} {\bibinfo {author} {\bibfnamefont {N.~P.}\ \bibnamefont
  {Ong}},\ }\href {\doibase 10.1103/PhysRevB.43.193} {\bibfield  {journal}
  {\bibinfo  {journal} {Phys. Rev. B}\ }\textbf {\bibinfo {volume} {43}},\
  \bibinfo {pages} {193} (\bibinfo {year} {1991})}\BibitemShut {NoStop}%
\bibitem [{\citenamefont {Podolsky}\ and\ \citenamefont {Kim}(2011)}]{daniel}%
  \BibitemOpen
  \bibfield  {author} {\bibinfo {author} {\bibfnamefont {D.}~\bibnamefont
  {Podolsky}}\ and\ \bibinfo {author} {\bibfnamefont {Y.~B.}\ \bibnamefont
  {Kim}},\ }\href {\doibase 10.1103/PhysRevB.83.054401} {\bibfield  {journal}
  {\bibinfo  {journal} {Phys. Rev. B}\ }\textbf {\bibinfo {volume} {83}},\
  \bibinfo {pages} {054401} (\bibinfo {year} {2011})}\BibitemShut {NoStop}%
\bibitem [{\citenamefont {Qi}\ \emph {et~al.}(2012)\citenamefont {Qi},
  \citenamefont {Korneta}, \citenamefont {Wan}, \citenamefont {DeLong},
  \citenamefont {Schlottmann},\ and\ \citenamefont {Cao}}]{cao-bi227-1}%
  \BibitemOpen
  \bibfield  {author} {\bibinfo {author} {\bibfnamefont {T.~F.}\ \bibnamefont
  {Qi}}, \bibinfo {author} {\bibfnamefont {O.~B.}\ \bibnamefont {Korneta}},
  \bibinfo {author} {\bibfnamefont {X.}~\bibnamefont {Wan}}, \bibinfo {author}
  {\bibfnamefont {L.~E.}\ \bibnamefont {DeLong}}, \bibinfo {author}
  {\bibfnamefont {P.}~\bibnamefont {Schlottmann}}, \ and\ \bibinfo {author}
  {\bibfnamefont {G.}~\bibnamefont {Cao}},\ }\href
  {http://stacks.iop.org/0953-8984/24/i=34/a=345601} {\bibfield  {journal}
  {\bibinfo  {journal} {Journal of Physics: Condensed Matter}\ }\textbf
  {\bibinfo {volume} {24}},\ \bibinfo {pages} {345601} (\bibinfo {year}
  {2012})}\BibitemShut {NoStop}%
\bibitem [{\citenamefont {{Baker}}\ \emph {et~al.}(2013)\citenamefont
  {{Baker}}, \citenamefont {{Moeller}}, \citenamefont {{Pratt}}, \citenamefont
  {{Hayes}}, \citenamefont {{Blundell}}, \citenamefont {{Lancaster}},
  \citenamefont {{Qi}},\ and\ \citenamefont {{Cao}}}]{cao-bi227-2}%
  \BibitemOpen
  \bibfield  {author} {\bibinfo {author} {\bibfnamefont {P.~J.}\ \bibnamefont
  {{Baker}}}, \bibinfo {author} {\bibfnamefont {J.~S.}\ \bibnamefont
  {{Moeller}}}, \bibinfo {author} {\bibfnamefont {F.~L.}\ \bibnamefont
  {{Pratt}}}, \bibinfo {author} {\bibfnamefont {W.}~\bibnamefont {{Hayes}}},
  \bibinfo {author} {\bibfnamefont {S.~J.}\ \bibnamefont {{Blundell}}},
  \bibinfo {author} {\bibfnamefont {T.}~\bibnamefont {{Lancaster}}}, \bibinfo
  {author} {\bibfnamefont {T.~F.}\ \bibnamefont {{Qi}}}, \ and\ \bibinfo
  {author} {\bibfnamefont {G.}~\bibnamefont {{Cao}}},\ }\href@noop {}
  {\bibfield  {journal} {\bibinfo  {journal} {ArXiv e-prints}\ } (\bibinfo
  {year} {2013})},\ \Eprint {http://arxiv.org/abs/1302.6905} {arXiv:1302.6905
  [cond-mat.str-el]} \BibitemShut {NoStop}%
\end{thebibliography}%
\end{document}